\documentclass[draftclsnofoot,journal,onecolumn]{IEEEtran}
\IEEEoverridecommandlockouts

\usepackage{ifthen}

\usepackage[tbtags]{amsmath} 
\usepackage{amssymb}  
\usepackage{verbatim} 
\usepackage{amsxtra}  

\usepackage{mathabx}

\usepackage{enumerate}

\usepackage{units}

\usepackage{amsfonts}
\usepackage{color}

\usepackage{cuted}

\providecommand{\nth}[1]{#1^\mathrm{th}}

\usepackage{subcaption}

\usepackage{wasysym}

\usepackage{cite}

\usepackage{mathtools}

\usepackage{MnSymbol}

\newcommand{\isi}[1]{}
\newcommand{\isiincl}[2]{}

\newcommand{\googleincl}[2]{}
\newcommand{\googleinclabs}[3]{}

\usepackage{tikz}
\usetikzlibrary{shapes,arrows,decorations.markings,positioning}

\tikzstyle{plant} = [draw, fill=red!5, rectangle, 
    minimum height=3em, minimum width=6em]
\tikzstyle{block} = [draw, fill=blue!5, rectangle, 
    minimum height=3em, minimum width=6em]
\tikzstyle{sum} = [draw, fill=yellow!10, circle, node distance=1cm]
\tikzstyle{coord} = [coordinate]
\tikzstyle{gain} = [draw, fill=red!5, regular polygon, regular polygon sides=3, shape border rotate=-90]
\tikzstyle{pinstyle} = [pin edge={to-,thick,black}]

\tikzstyle{BitPipe} = [thick, decoration={markings,mark=at position
   1 with {\arrow[semithick]{open triangle 60}}},
   double distance=1.4pt, shorten >= 5.5pt,
   preaction = {decorate},
   postaction = {draw,line width=1.4pt, white,shorten >= 4.5pt}]

\usepackage{epsfig, graphicx, psfrag}

\usepackage[mathcal]{euscript}
\usepackage[cal=boondox]{mathalfa}

\usepackage{amsthm}

\newtheorem{thm}{Theorem}

\newtheorem{lem}{Lemma}

\newtheorem{corol}{Corollary}

\theoremstyle{definition}
\newtheorem{defn}{Definition}
\newtheorem*{defn*}{Definition}

\newtheorem*{scheme*}{Scheme}

\newtheorem{example}{Example}

\theoremstyle{remark}
\newtheorem{remark}{Remark}

\providecommand{\thmref}[1]{Th.~\ref{#1}}

\providecommand{\defnref}[1]{Def.~\ref{#1}}
\providecommand{\secref}[1]{Sec.~\ref{#1}}
\providecommand{\lemref}[1]{Lem.~\ref{#1}}

\providecommand{\figref}[1]{Fig.~\ref{#1}}
\providecommand{\colref}[1]{Corol.~\ref{#1}}
\providecommand{\corolref}[1]{Corol.~\ref{#1}}
\providecommand{\appref}[1]{App.~\ref{#1}}

\usepackage{hyperref}

\newcommand{\ie}{i.e.}

\newcommand{\eg}{e.g.}

\newcommand{\reals}{\mathbb{R}}
\newcommand{\ints}{\mathbb{Z}}
\newcommand{\compls}{\mathbb{C}}
\newcommand{\nats}{\mathbb{N}}

\providecommand{\trace}[1]{\mathrm{trace}\left\{ #1 \right\}}

\newcommand{\bm}[1]{\mbox{\boldmath{$#1$}}}

\newcommand{\e}{\mathrm{e}}

\newcommand{\mS}{\mathcal{S}}

\newcommand{\Comment}[1]{}
\newcommand{\old}[1]{}
\newcommand{\rem}[1]{}

\newcommand{\tb}{\tilde b}

\newcommand{\ba}{\bm{a}}
\newcommand{\bba}{\bar{\ba}}
\newcommand{\bbba}{\bar{\bba}}
\newcommand{\bA}{\text{\bf A}}
\newcommand{\bc}{\bm{c}}
\newcommand{\bbc}{\bar{\bc}}
\newcommand{\tbc}{\widetilde{\bc}}
\newcommand{\bbbc}{\bar{\bbc}}
\newcommand{\bmu}{\bm \mu}
\newcommand{\bC}{\text{\bf C}}

\newcommand{\bI}{\text{\bf I}}
\newcommand{\bJ}{\text{\bf J}}
\newcommand{\bq}{\bm{q}}
\newcommand{\bQ}{\text{\bf Q}}

\providecommand{\bU}{\text{\bf U}}
\providecommand{\bV}{\text{\bf V}}

\newcommand{\bX}{{\bm X}}

\providecommand{\bu}{{\bm u}}
\providecommand{\bv}{{\bm v}}

\providecommand{\bV}{{\bf V}}

\newcommand{\cS}{{\mathcal S}}

\providecommand{\infinity}{\infty}

\newcommand{\abs}[1]{\left| #1 \right|}
\newcommand{\Norm}[1]{\left\| #1 \right\|}

\providecommand{\e}{{\rm e}}
\providecommand{\comment}[1]{}

\providecommand{\norm}[1]{\Norm{#1}}

\newcommand{\beqn}[1]{\begin{eqnarray}\label{#1}}
\newcommand{\eeqn}{\end{eqnarray}}
\newcommand{\beq}[1]{\begin{equation}\label{#1}}
\newcommand{\eeq}{\end{equation}}

\providecommand{\var}[1]{{\mathrm{Var}\left\{ #1 \right\}}}

\providecommand{\trace}{\text{trace}}

\providecommand{\half}{\frac{1}{2}}

\providecommand{\trace}{\text{trace}}

\providecommand{\half}{\frac{1}{2}}

\makeatletter
\newcommand{\vast}{\bBigg@{4}}
\newcommand{\Vast}{\bBigg@{5}}
\makeatother

\providecommand{\bbE}{\mathbb{E}}

\providecommand{\E}[1]{\bbE \left\{ #1 \right\}}

\providecommand{\LMMSE}{\overline{\mathcal{E}^2}}
\providecommand{\LMMSEone}{\overleftarrow{\mathcal{E}^2}}
\providecommand{\LMMSEtwo}{\overleftrightarrow{\mathcal{E}^2}}
\providecommand{\LMMSEtwoP}[1]{\overleftrightarrow{\mathcal{E}^2_{#1}}}
\providecommand{\Xothers}[2]{\bX_{[#2 \backslash #1]}}
\providecommand{\Cn}[1]{\bC_{#1}}
\providecommand{\Cvec}[1]{\bm{C}_{#1}}
\providecommand{\tCn}[1]{\tilde{\bC}_{#1}}
\providecommand{\Cothers}[2]{\bC_{[#2 \backslash #1]}}
\providecommand{\CwithOthers}[2]{\bm{C}_{#1,[#2 \backslash #1]}}
\providecommand{\bzero}{{\bf 0}}
\providecommand{\veczero}{\text{\textbf{\textit{0}}}}
\providecommand{\except}[2]{\left[ #2 \middle\backslash #1 \right]}

\DeclareMathOperator{\sign}{sign}
\providecommand{\MaxTSP}{\mathrm{MaxTSP}}
\providecommand{\MinTin}{\mathrm{MinTin}}

\mathtoolsset{showonlyrefs=true}

\usepackage{comment}

\begin{document}
\title{Monotonicity of the Trace--Inverse of Covariance Submatrices and \\ Two-Sided Prediction}

\author{\textit{At present, the future is just as important as the past.} \\ \ \\
Anatoly Khina, Arie Yeredor, and Ram Zamir
	\thanks{This research was supported by the \textsc{Israel Science Foundation} (grants No.\ 2077/20, 2427/19, and No.\ 2623/20).
	The work of A.~Khina was supported by the WIN Consortium through the Israel Ministry of Economy and Industry.}
	\thanks{The authors are with the Department of Electrical Engineering--Systems, Tel Aviv University, Tel Aviv, Israel 6997801. \mbox{E-mails}: \mbox{{\tt \{anatolyk,arie,zamir\}@eng.tau.ac.il}}}
}

\maketitle
	
	
\begin{abstract}
    It is common to assess the ``memory strength'' of a stationary process
    by looking at how fast the normalized log--determinant of its
    covariance submatrices (\ie, entropy rate) decreases.
    In this work, we propose an alternative characterization in terms
    of the normalized trace--inverse of the covariance submatrices.
    We show that this sequence is monotonically non-decreasing and
    is constant if and only if the process is white. Furthermore,
    while the entropy rate is associated with one-sided prediction errors
    (present from past), the new measure is associated with two-sided
    prediction errors (present from past and future).
    This measure can be used as an alternative to Burg's maximum-entropy principle
    for spectral estimation.
    We also propose 
    a counterpart for 
    non-stationary processes, by looking
    at the average trace--inverse of subsets.
\end{abstract}

\begin{IEEEkeywords}
    Maximum entropy, prediction, minimum mean square error, causality.
\end{IEEEkeywords}

\allowdisplaybreaks

\section{Introduction}
\label{s:intro}

The entropy rate of a stationary process $\{X_1, X_2, X_3, \ldots\}$ is given by the limit
$\lim_{n \to \infty} H_n$
of the normalized joint differential entropy $h(\cdot)$ of $n$ consecutive samples:
\begin{align} 
\label{eq:average-entropy}
    H_n \triangleq \frac{1}{n} h(X_1,...,X_n).
\end{align}

It is well known that $H_n$ is monotonically non-increasing in $n$,
and is constant if and only if (iff) the process is memoryless [has independent and identically distributed (i.i.d.) samples] \cite[Sec.~II-B]{DemboCoverThomas:inequalities:IT91}, \cite[Ch.~17.6]{CoverBook2Edition}.
Furthermore, the difference
\begin{align} 
    D_n \triangleq H_1 - H_n
\end{align}
is equal to the Kullback--Leibler (KL) divergence \cite[Ch.~2.3]{CoverBook2Edition} between the $n$-dimensional
distribution of the process and the $n$-dimensional distribution of a memoryless process with the same marginal. If the process is Gaussian, then $D_n$ simplifies to \cite[Ch.~8.4]{CoverBook2Edition}
\begin{align}
\label{eq:log-det-ratio}
    D_n^G = \half \log \frac{\var{X_1}}{\abs{\Cn{n}}^{1/n}} \:,
\end{align}
where $\Cn{n}$ is the $\nth{n}$ order covariance matrix of the process, $\abs{\Cn{n}}$ denotes its determinant, and $\var{X_1}$ is the variance of $X_1$ (and of any other sample, by stationarity).
This non-negative quantity is zero iff the $\nth{n}$ order covariance $\Cn{n}$ is proportional to the identity matrix $\bI_n$,
i.e., the vector $(X_1, \ldots,X_n)$ is white.
Thus, $D_n^G$ can be thought of as a measure of the memory strength (``distance from whiteness'') which increases with $n$ for a process with memory \cite[Th.~27]{DemboCoverThomas:inequalities:IT91}, \cite[Th.~17.9.6]{CoverBook2Edition}, as implied by the monotonicity of $H_n$ \eqref{eq:average-entropy}.

The quantity $D_n^G$ has two additional interesting interpretations:
The first follows from writing the trace and the determinant of $\Cn{n}$ as the sum and product, respectively,
of its eigenvalues $\lambda_1,...,\lambda_n$:
\begin{align} 
\label{eq:arithmetic-to-geometric}
    D_n^G = \half \log \frac{\frac{1}{n} \sum_{i=1}^n \lambda_i }{\left( \prod_{i=1}^n \lambda_i \right)^{1/n}} \:,
\end{align}
\ie, it is half the logarithm of the arithmetic-to-geometric means ratio of the eigenvalues, which is zero iff the eigenvalues are equal by the arithmetic mean--geometric mean inequality (recall that the eigenvalues of a covariance matrix are real and non-negative).
The second follows from the chain rule for entropies, \ie, from writing
the joint differential entropy as the sum of conditional differential entropies:
\begin{align} 
    n H_n \!= h(X_1) + h(X_2|X_1) + \cdots + h(X_n|X_{n-1},X_{n-2},\ldots, X_1),
\nonumber
\end{align}
where, in the Gaussian case, the terms are associated with prediction minimum mean square errors (MMSEs) of
increasing order.
Thus, $D_n^G$ may be viewed as the mean prediction gain 
\begin{align} 
\label{eq:one-sided-prediction}
    D_n^G = \sum_{i=1}^n \half \log \frac{\var{X_i}}{\LMMSE(i|\{i-1, i-2, \ldots, 1\})} ,
\end{align}
where $\LMMSE(i|\{i-1, i-2, \ldots, 1\})$ is the $\nth{i}$ order linear prediction MMSE of $X_i$ given $\{X_{i-1}, X_{i-2}, \ldots, X_1\}$,
and, therefore, $D_n^G$ is zero iff the prediction MMSEs are all equal to $\var{X_1}$.

Jaynes' maximum entropy (MaxEnt) principle \cite{Jaynes57} and its specialization to spectral estimation by Burg \cite{Burg:PhD1975:MaximumEntropy}, \cite[Ch.~12.6]{CoverBook2Edition}
dictate that, for given $k+1$ consecutive autocorrelation constraints 
\begin{align}
\label{eq:correlation-constraints}
    \E{X_{i+\ell} X_i} &= c_\ell , & \ell \in \{ 0, 1, 2, \ldots, k \},
\end{align}
a zero-mean Gaussian autoregressive (AR) process of order $k$, AR($k$), maximizes the entropy $H_n$ for all $n \geq k$.
This process, thus, minimizes the memory strength (``distance from whiteness'') $D_n^G$ under the given $k+1$ correlation constraints~\eqref{eq:correlation-constraints}.

In this work, an alternative measure of the memory strength 
of a stationary process---the trace of the inverse of a covariance matrix (also referred to as the \textit{precision matrix} \cite[Lib.~II, Sec.~III]{Gauss:Astronomy:Book:1809}) is considered.
The \textit{trace--inverse} (Tin) is a common measure of confidence in data reliability analysis (see, \eg, \cite{KalantzisBekasCurioniGallopoulos:NumAlg2013:Tin} and references therein), where it is usually sought to be {\em maximized}; various approaches have been proposed for bounding \cite{BaiGolub:NumMath1996:Tin_Det} or for explicitly estimating \cite{Meurant:NumAlg2009:Tin} or calculating \cite{BrezinskiFikaMitrouli:NumLinAlgApps2012:Tin} this quantity in different contexts (albeit, not in the context of stationary processes). 

By borrowing and employing this measure for consecutive samples of discrete-time 
stochastic stationary processes, we unveil interesting dual relationships to the ones above,
where the Tin plays the role of the log--determinant \eqref{eq:log-det-ratio},
arithmetic-to-harmonic means ratio replaces the arithmetic-to-geometric means ratio \eqref{eq:arithmetic-to-geometric},
and two-sided prediction (TSP) \cite[Ch.~I.10]{Rozanov:Book1967}, \cite{HelseyGriffiths:TSP:ICASSP1976,Kay:TSP:TSP1983,Picinbono:TSP:InFrench,Picinbono:TSP:TSP1988,HsueYagle:TSPvsOSP:TSP1995} plays the role of the usual one-sided prediction (OSP) \eqref{eq:one-sided-prediction}.

In particular, the Tin of $n$ consecutive samples of a stationary process, normalized by $n$, is shown to be monotonically non-decreasing with $n$, and constant iff the process is white.
Furthermore, a new criterion for spectral estimation/completion is proposed, 
seeking the {\em minimization} of the Tin as an indication of maximized uncertainty (poor ``reliability”) of the process, reminiscent of the notion of entropy in this context. This minimum Tin (MinTin) criterion is shown to yield a different solution, in general, from the familiar MaxEnt (Gaussian AR) solution of Burg.

The rest of the paper is organized as follows. In the following subsection we introduce our notations. We state the main result of this work---the monotonicity of the normalized Tin----in \secref{s:stationary}. We then provide proofs of the main result using two different approaches which elucidate different insights---via MMSE estimation and via AR modeling, in Secs.\ \ref{s:proofs} and \ref{s:AR}, respectively. 
We introduce a MinTin criterion for spectral estimation/completion and contrast this criterion with the MaxEnt criterion in \secref{s:mintin-maxent}.
Finally, we derive parallel results for non-stationary processes in \secref{s:non-stationary}, and conclude the paper with a discussion in \secref{s:summary}.


\subsection{Notation}
\label{ss:notation}

$\mathbb{E}$, $\mathrm{Var}$, $(\cdot)^T$, and $(\cdot)^*$ denote the expectation, variance, transpose, and complex conjugation operations, respectively. 
$j = \sqrt{-1}$ is the imaginary~unit.

Denote the natural, integer, and real numbers by $\nats$, $\ints$, and $\reals$, respectively. Denote the set of the smallest $n \in \nats$ natural numbers by $[n] \triangleq \{1, \ldots, n\}$, 
and this set with 
$i \in \nats$ removed---by $\except{i}{n} \triangleq [n] \backslash \{i\}$, where `$\backslash$' denotes the set difference operation.

We denote matrices by uppercase boldface roman letters ($\bA$), and column vectors---by (lowercase or uppercase) boldface italic or Greek letters ($\bv$, $\bX$, $\bmu$). $\bI_n$ denotes the identity matrix of size $n$.

$\abs{\bA}$ and $\abs{\cS}$ denote the determinant of the matrix $\bA$ and the cardinality of the set $\cS$, respectively.

Denote $\bX_\cS = (X_{i_1}, X_{i_2}, \ldots, X_{i_{|\cS|}})^T$, where 
$\cS \subset \ints$, 
$i_\ell \in \mS$ for all $\ell \in [|\cS|]$ such that $i_1 < i_2 < \cdots < i_{|\cS|}$.\footnote{If $\cS$ is a countably infinite set, then the requirement $\ell \in [|\cS|]$ is replaced with $\ell \in \nats$ and no maximal index (paralleling $i_{|S|}$ in the finite-set $\cS$ case) exists.}

Denote by $\Cn{\cS} \triangleq \E{\bX_\cS \bX_\cS^T}$ the autocovariance matrix of a zero-mean random vector $\bX_\cS$ where $\cS \subset \ints$, $|\cS| < \infty$, 
and---by $\Cn{\cS_1,\cS_2} \triangleq \E{ \bX_{\cS_1} \bX_{\cS_2}^T }$ the cross-covariance matrix between $\bX_{\cS_1}$ and $\bX_{\cS_2}$, where $\cS_1, \cS_2 \subset \ints$, and $|\cS_1|, |\cS_2| < \infty$.

With some abuse of notation, we further denote $\Cn{n} \triangleq \Cn{[n]}$, $\Cvec{i,\cS} \triangleq \Cvec{\{i\}, \cS}$, and $C_{i,i} \triangleq C_{\{i\},\{i\}}$, where $n \in \nats$, $i \in \ints$, and $\cS \subset \ints$. Note that the matrix $\Cn{1}$ is a scalar and $\Cn{i, \cS}$ is a (row) vector, and therefore they will be denoted $C_1$ and $\Cvec{i, \cS}$, respectively.

The $(i,j)$ element of a matrix $\bA$ will be denoted by $\left[ \bA \right]_{i,j}$, and its principal upper-left $n \times n$ submatrix---by $\left[ \bA \right]_{[n],[n]}$.

Denote by 
$\LMMSE(i|\cS)$
the MMSE achievable by linear estimation (LMMSE) in estimating 
$X_i$ given $\bX_\cS$ for $i \in \ints$, $\cS \subset \ints$.
For a set $\cS \subset \ints$, $|\cS| < \infty$, this LMMSE is known to be given by \cite[Ch.~7.3]{PapoulisBook:4thEd}:
\begin{align}
\label{eq:LMMSE}
    \LMMSE(i|\cS) = C_{i,i} - \Cvec{i,\cS} \Cn{\cS}^{-1} \Cvec{i,\cS}^T \:.
\end{align}

For a zero-mean, wide-sense stationary (WSS) process $\{X_k|k\in\ints\}$, we further define the following: 
\begin{itemize}
    \item The covariance sequence
    \begin{align}
        c_\ell \triangleq \E{X_{i+\ell}X_i} \;\;\; \forall \ell\in\ints.
    \end{align}
    In particular, $c_0$ denotes the variance of the process;
    \item The power spectral density (spectrum)
    \begin{align}
        S \left( e^{j 2\pi f} \right) \triangleq 
        \sum_{\ell = -\infty}^\infty
        c_\ell \e^{-j 2 \pi f \ell} 
    \end{align}
    (assuming convergence of the sum for all $f\in\reals$).
    \item The OSP LMMSE---the LMMSE in ``predicting" the present given the entire past---
    \begin{align}
    \label{eq:one-sided-LMMSE}
        \LMMSEone \triangleq \LMMSE(i|\{ \ell < i | \ell \in \ints \}) = \LMMSE(i| i-1, i-2,\ldots ); 
    \end{align}
    The TSP LMMSE---the LMMSE in predicting the present given the entire past and future---
\end{itemize}
    \begin{align}
        \LMMSEtwo \triangleq \LMMSE(i|\ints \backslash \{i\}) = \LMMSE(i| \ldots, i-2, i-1, i+1, i+2, \ldots ).\ 
    \label{eq:two-sided-LMMSE}
    \end{align}
Note that, due to the stationarity of the process, both the OSP and the TSP LMMSEs do not depend on $i$; similarly, $\Cn{n}$ denotes the covariance of {\em any} $n$ consecutive samples of the process.

The (real-valued) eigenvalues of an $n \times n$ symmetric matrix $\bA$, 
non-increasingly ordered,
will be denoted by 
\begin{align} 
    \lambda^\downarrow_1(\bA) \geq \lambda^\downarrow_2(\bA) \geq \cdots \geq \lambda^\downarrow_n(\bA) .
\end{align}

We use the conventions $1/0 \triangleq \infty$, and---by extension---$\trace{\bA^{-1}} \triangleq \infty$ for a singular PSD matrix $\bA$.


\section{Trace--Inverse for Stationary Processes}
\label{s:stationary}

Let $\{ X_k | k \in \ints \}$ be a zero-mean WSS process. We define the normalized trace--inverse of $\Cn{n}$ (the covariance matrix of any $n$ consecutive samples) as follows.

\begin{defn}[Normalized trace--inverse]
\label{def:trace-inverse}
    The normalized Tin of order $n$ is defined as
    \begin{align}
        M_n \triangleq \frac{1}{n} \trace{\Cn{n}^{-1}}.
    \end{align}
\end{defn}

We are now ready to state the main result of this work; two alternative proofs thereof (providing different insights, as explained in the sequel) are provided in Secs.~\ref{s:proofs} and \ref{s:AR}, using MMSE estimation tools and AR modeling, respectively.
\begin{thm}[Normalized-Tin monotonicity]
\label{thm:trace-inverse:monotone}
    The sequence $\{M_n | n \in \nats \}$ is monotonically 
    non-decreasing. Namely, for all $n\in\nats$, 
    \begin{align}
    \label{eq:trace-inverse:monotone}
        M_n &\leq M_{n+1},    \end{align}
    with equality in \eqref{eq:trace-inverse:monotone} iff
    either $\Cn{n+1} = c_0 \cdot \bI_{n+1}$, in which case $M_1 = M_2 = \cdots = M_{n+1} = 1/c_0$ (note that an equality for all $n\in\nats$ holds iff $\{X_k\}$ is white);  
    or $\Cn{n}$ is singular, in which case $M_n = M_{n+1} = \infty$.
\end{thm}

By \thmref{thm:trace-inverse:monotone}, the sequence $\{M_n | n \in \nats\}$ is monotonically non-decreasing, meaning that its minimum and maximum are $M_1 = 1 / c_0$ and $M_\infty \triangleq \lim_{n \to \infty} M_n$, respectively, where we shall see in \secref{s:proofs} that the latter equals the reciprocal of the MMSE achievable by linear prediction of the present given the entire future and past.

\section{Proof of \thmref{thm:trace-inverse:monotone} via MMSE Estimation}
\label{s:proofs}

In this section, we prove \thmref{thm:trace-inverse:monotone} using MMSE-estimation tools and matrix properties by reinterpreting the diagonal elements of $\Cn{n}^{-1}$ as the reciprocals of the LMMSEs in estimating the different $X_i$'s given the rest.

The following are well known results and can be found, \eg, in \cite[Ch.~0.7]{HornJohnsonBook:MatrixAnalysis}.

\begin{lem}[Partitioned-matrix inversion]
\label{lem:InverseMatrix:Schur}
    Let $\bA$ be an invertible matrix of dimensions $n \times n$, partitioned as follows.
    \begin{align}
        \bA = 
        \begin{bmatrix}
            \bA_{11} & \bA_{12}
         \\ \bA_{21} & \bA_{22}
        \end{bmatrix}
        ,
    \end{align}
    where the submatrix $\bA_{ij}$ is of dimensions $n_i \times n_j$ for $i,j \in [2]$, and $n_i \in [n]$ such that $n_1 + n_2 = n$. Assume further that $\bA_{11}$ and $\bA_{22}$ are invertible. Then, 
    \begin{align}
        \bA^{-1} = 
        \begin{bmatrix}
            \left( \bA \backslash \bA_{22} \right)^{-1} & -\bA_{11}^{-1} \bA_{12} \left( \bA \backslash \bA_{11} \right)^{-1}
         \\ -\bA_{22}^{-1} \bA_{21} \left( \bA \backslash \bA_{22} \right)^{-1} & \left( \bA \backslash \bA_{11} \right)^{-1}
        \end{bmatrix}
        ,
    \end{align}
    where $\bA \backslash \bA_{ii} \triangleq \bA_{jj} - \bA_{ji} \bA_{ii}^{-1} \bA_{ij}$ is the Schur complement of $\bA_{ii}$ in $\bA$  \cite[Ch.~0.8]{HornJohnsonBook:MatrixAnalysis} for $i,j \in [2]$ and $i \neq j$,
    and its inverse is further equal, by the matrix inversion lemma, to
    \begin{align}
    \label{eq:matrix-inversion-lemma}
    \begin{aligned}
        \left( \bA \backslash \bA_{ii} \right)^{-1} &=
        \left( \bA_{jj} - \bA_{ji} \bA_{ii}^{-1} \bA_{ij} \right)^{-1}\\
        &=\bA_{jj}^{-1} + \bA_{jj}^{-1}\bA_{ji}\left( \bA\backslash\bA_{jj} \right)^{-1} \bA_{ij}\bA_{jj}^{-1}.
    \end{aligned}
    \end{align}
\end{lem}

The following result is due to Hesley and Griffiths \cite{HelseyGriffiths:TSP:ICASSP1976}, first proved by Kay, using Lagrange multipliers \cite{Kay:TSP:TSP1983}. We provide an alternative proof for completeness, using \lemref{lem:InverseMatrix:Schur}.
\begin{lem}[Inverse-covariance diagonal via LMMSE]
\label{lem:trace-LMMSEs}
    Assume $\Cn{n}$ is invertible. Then, the diagonal elements of its inverse are given by the reciprocals of the LMMSEs in estimating the corresponding $X_i$'s given the rest:
    \begin{align}
    \label{eq:diagonal}
        \left[ \Cn{n}^{-1} \right]_{i,i} &= \frac{1}{\LMMSE(i | \except{i}{n})}, 
        & i &\in [n].
    \end{align}
\end{lem}

\textit{\, \, Proof:}
    The first diagonal element of $\Cn{n}^{-1}$ is equal to 
    \begin{subequations}
    \label{eq:diagonal_first}
    \begin{align}
        \left[ \Cn{n}^{-1} \right]_{1,1} &= \left( C_{1,1} - \CwithOthers{1}{n} \Cothers{1}{n}^{-1} \CwithOthers{1}{n}^T \right)^{-1}
    \label{eq:diagonal_first:Schur}
     \\ &= \frac{1}{\LMMSE(1 | \except{1}{n})},
    \label{eq:diagonal_first:LMMSE}
    \end{align}
    \end{subequations}
    where \eqref{eq:diagonal_first:Schur} holds by \lemref{lem:InverseMatrix:Schur}
    and \eqref{eq:diagonal_first:LMMSE} follows from \eqref{eq:LMMSE}.
    By rearranging the entries of $\bX_{[n]}$, one arrives at
    a similar expression for the $i^\mathrm{th}$ ($i \in [n]$) diagonal element of $\Cn{n}^{-1}$:
    \begin{align}
        \left[ \Cn{n}^{-1} \right]_{i,i} &= \left( C_{i,i} - \CwithOthers{i}{n} \Cothers{i}{n}^{-1} \CwithOthers{i}{n}^T \right)^{-1}
     \\ &= \frac{1}{\LMMSE(i | \except{i}{n})} .
    \tag*{\IEEEQED}
    \end{align}

\begin{corol}[Tin via LMMSEs]
\label{col:Tin-LMMSEs}
    $M_n$, the normalized Tin of 
    order $n$, 
    is equal to 
    the mean 
    of the reciprocals of the LMMSEs of the corresponding $X_i$'s given the rest:
    \begin{align}
    \label{eq:trace-LMMSE}
        M_n = \frac{1}{n} \sum_{i=1}^n \frac{1}{\LMMSE(i | \except{i}{n})} \,.
    \end{align}
\end{corol}

\begin{IEEEproof}
    If $\Cn{n}$ is singular,
    there exists $i \in [n]$ such that $X_i$ is a linear combination of $\Xothers{i}{n}$. Thus, $\LMMSE(i|\except{i}{n}) = 0$ and $M_n = \infty$ by our conventions (recall \secref{ss:notation}).
    
    For an invertible $\Cn{n}$, the result 
    follows from \lemref{lem:trace-LMMSEs} by summing all the diagonal elements \eqref{eq:diagonal} and dividing by $n$. 
\end{IEEEproof}

The following lemma is an immediate consequence of the stationarity of the process and the fact that adding more measurements cannot increase the LMMSE.
\begin{lem}[LMMSE monotonicity]
\label{lem:LMMSE_more_measurments}
    The following relations hold for all $n \in \nats$ and $i \in [n]$:
    \begin{subequations}
    \label{eq:LMMSE-inequality}
    \noeqref{eq:LMMSE-inequality:itoi,eq:LMMSE-inequality:itoi+1}
    \begin{align}
        \LMMSE(i | \except{i}{n}) &\geq \LMMSE(i | \except{i}{n+1}) ,
    \label{eq:LMMSE-inequality:itoi}
     \\ \LMMSE(i | \except{i}{n}) &\geq \LMMSE(i+1 | \except{i+1}{n+1}) . \quad
    \label{eq:LMMSE-inequality:itoi+1}
    \end{align}
    \end{subequations}
\end{lem}

\begin{IEEEproof}
        In \eqref{eq:LMMSE-inequality:itoi}, $X_{n+1}$ is added to the set $\bX_{\except{i}{n}}$, whereas in \eqref{eq:LMMSE-inequality:itoi+1}, $X_0$ is added to the same set, followed by an index-shift, which is immaterial due to stationarity. 
\end{IEEEproof}

We are now ready to prove \thmref{thm:trace-inverse:monotone}.

\begin{IEEEproof}[Proof of \thmref{thm:trace-inverse:monotone}]
    \underline{\textit{Invertible covariance matrices:}} Assume that $\Cn{n}$ and $\Cn{n+1}$ are invertible.
    Since at least one of the elements of a (non-empty) set is at least as large as its mean, 
    there exists an index 
    $\ell \in [n]$ such that
    \begin{subequations}
    \label{eq:LMMSE-mean}
    \noeqref{eq:LMMSE-mean:Mn}
    \begin{align}
        \frac{1}{\LMMSE(\ell|\except{\ell}{n})} &\geq \frac{1}{n} \sum_{i=1}^n \frac{1}{\LMMSE(i|\except{i}{n})}
    \label{eq:LMMSE-mean:LMMSEs}
     \\ &= M_n \,.
    \label{eq:LMMSE-mean:Mn}
    \end{align}
    \end{subequations}
    
    This allows us to arrive at the desired result:
    \begin{subequations}
    \noeqref{eq:traces-inequality:break}
    \label{eq:traces-inequality}
    \begin{align}
        (n+1) M_{n+1} &= \sum_{i=1}^{n+1} \frac{1}{\LMMSE(i|\except{i}{n+1})}
    \label{eq:traces-inequality:trace-LMMSE-n+1}
     \\ &= \sum_{i=1}^{\ell-1} \frac{1}{\LMMSE(i|\except{i}{n+1})} + \frac{1}{\LMMSE(\ell | \except{\ell}{n+1})} 
     + \sum_{i=\ell+1}^{n+1} \frac{1}{\LMMSE(i|\except{i}{n+1})} \ \ 
    \label{eq:traces-inequality:break}
     \\ &\geq \sum_{i=1}^{\ell-1} \frac{1}{\LMMSE(i | \except{i}{n})} + M_n + \sum_{i=\ell}^n \frac{1}{\LMMSE(i | \except{i}{n})} \quad\ \ 
    \label{eq:traces-inequality:prev-ineqs}
     \\ &= (n+1) M_n ,
    \label{eq:traces-inequality:trace-LMMSE-n}
    \end{align}
    \end{subequations}
    where \eqref{eq:traces-inequality:trace-LMMSE-n+1} and \eqref{eq:traces-inequality:trace-LMMSE-n} are due to \lemref{lem:trace-LMMSEs}, 
    and \eqref{eq:traces-inequality:prev-ineqs} follows from \lemref{lem:LMMSE_more_measurments} and \eqref{eq:LMMSE-mean}.
    
    \underline{\textit{Singular covariance matrices:}} 
    If $\Cn{n}$ is singular, then, by \lemref{lem:trace-LMMSEs} and our convention (recall \secref{ss:notation}), $M_n = \infty$, and therefore, by \lemref{lem:LMMSE_more_measurments}, $M_{n+1} = \infty$, and \eqref{eq:trace-inverse:monotone} holds with equality.
    If $\Cn{n}$ is invertible but $\Cn{n+1}$ is not, then, by \lemref{lem:trace-LMMSEs}, $M_n < \infty$ but $M_{n+1} = \infty$, and the inequality \eqref{eq:trace-inverse:monotone} strictly holds.

    \underline{\textit{Equality condition:}}
    The proof of the equality condition is technical and is therefore relegated to 
    \appref{app:tin-monotonicity:equality-cond}.
    An alternative short proof is provided in \secref{s:AR}.
\end{IEEEproof}


\section{Proof of \thmref{thm:trace-inverse:monotone} via Autoregressive Modeling}
\label{s:AR}

In this section, we provide additional insights by proving \thmref{thm:trace-inverse:monotone} using results for 
autoregressive (AR) processes that enable to obtain explicit expressions for $M_n$. As we shall see, although the process $\{X_i\}$ may not be an AR process, in general, 
there exists an AR process of order up to $n$ that is consistent with \textit{any} admissible covariance matrix $\Cn{n+1}$ by virtue of the Yule--Walker equations \cite[Ch.~12]{PapoulisBook:4thEd};
thus, it suffices to prove \thmref{thm:trace-inverse:monotone} for this AR process. 

Consider a zero-mean stationary AR process $\{X_i\}$, \ie, a process that
can be represented as 
\begin{subequations}
\noeqref{eq:AR:model:compact,eq:AR:model:standard}
\label{eq:AR:model}
\begin{align}
\label{eq:AR:model:compact}
    \sum_{\ell=0}^p a_\ell X_{i-\ell} &= W_i & \forall i\in\ints,
\end{align}
or, equivalently, as 
\begin{align}
\label{eq:AR:model:standard}
    X_i &= - \sum_{\ell=1}^p a_\ell X_{i-\ell} + W_i & \forall i\in\ints,
\end{align}
\end{subequations}
where $\{W_i\}$ is a sequence of uncorrelated 
random variables with zero-mean and
variance $\sigma_W^2$,
$p \in \nats$ is the order of the AR process assuming $a_p \neq 0$, $a_0 = 1$, and $\{a_\ell | a_\ell \in \reals, \ell \in [p] \}$ are the process coefficients; the process is assumed stable, \ie, the poles of the polynomial 
\begin{align}
    A(z) = \sum_{\ell = 0}^p a_\ell z^{-\ell} 
\end{align}
are assumed to lie strictly inside the unit circle \cite[Ch.~10]{Doob:StochasticProcesses:Book}. Such a process is denoted by AR($p$).

All elements of $\Cn{n}^{-1}$ can be explicitly represented in terms of the process coefficients $\{a_i\}$ for $p \leq n$, 
as was proved in  \cite{Siddiqui:AR:inverse_covariance,Galbraith:AR:inverse_covariance} (see also \cite{Wise:AR:inverse_covariance,Champernowne:AR:inverse_covariance}). 
We next provide an alternative proof that relies on the Gohberg--Semen\c{c}ul formula \cite{GohbergSemencul:ToeplitzInverse} for the inverse of the covariance matrix of an AR process.

\begin{thm}[Inverse covariance of AR models]
\label{thm:AR:inverse_covariance}
    Assume a WSS AR process \eqref{eq:AR:model} of order $p \in [n]$.
    Then, the elements of its inverse autocovariance matrix $\Cn{n}^{-1}$ 
    are given as
    \begin{align}
        \left[ \Cn{n}^{-1} \right]_{i,j} \sigma_W^2 &= \sum_{\ell=0}^{i-1} a_\ell a_{\ell + j - i} - \sum_{\ell=n + 1 - j}^{n + i - j} a_\ell a_{\ell + j - i},
        & 1 &\leq i \leq j \leq n; \quad
    \nonumber
     \\ \left[ \Cn{n}^{-1} \right]_{j,i} &= \left[ \Cn{n}^{-1} \right]_{i,j}, & i,j &\in [n];
    \end{align}
    where $a_\ell = 0$ for $p < \ell \leq n$.
\end{thm}

\begin{IEEEproof}
    We prove the theorem for $p < n$ in here, and relegate the extension for $p = n$ to 
    \appref{app:inverse-covariance:p=n}.

    Denote by $\bv$ the first column of $\Cn{n}^{-1}$, \ie, 
    the vector $\bv \triangleq \begin{bmatrix} \bv_0 & \bv_1 & \cdots & \bv_{n-1} \end{bmatrix}^T$ that satisfies 
    \begin{align}
    \label{eq:Gohberg_semencul:first_col}
        \Cn{n} \bv = \begin{bmatrix} 1 & 0 & \cdots & 0 \end{bmatrix}^T . 
    \end{align}
    Note that $v_0 = |\Cn{n-1}|/|\Cn{n}|$ by Cramer's rule \cite[Ch.~0.8.3]{HornJohnsonBook:MatrixAnalysis}.
    Thus, $v_0 \neq 0$
    since $\Cn{n}$ and therefore also $\Cn{n-1}$ are invertible. 
    
    Then, according to the Gohberg--Semen\c{c}ul formula \cite{GohbergSemencul:ToeplitzInverse} 
    for symmetric matrices \cite{KailathVieiraMorf:InverseToeplitz,KailathBrucksteinMorgan:InverseToeplitz,MukherjeeMaiti:ToeplitzInverse},
    $\Cn{n}^{-1}$ is equal to 
    \begin{align}
    \label{eq:Gohberg-Semencul}
        \Cn{n}^{-1} = \frac{1}{v_0} \left( \bV \bV^T - \bU \bU^T \right), 
    \end{align}
    where $\bV$ and $\bU$ are lower-triangular Toeplitz matrices of dimensions $n \times n$, 
    $\bv$ is the first column of $\bV$, and 
    $\bu \triangleq \begin{bmatrix} 0 & v_{n-1} & v_{n-2} & \cdots & v_{1} \end{bmatrix}^T$ 
    is the first column of $\bU$.

    Now note that since the process is a WSS AR process, it satisfies the Yule--Walker equations \cite[Ch.~12]{PapoulisBook:4thEd}
    \begin{align}
    \label{eq:YuleWalker}
    \!\!\!\!
        \Cn{n} \begin{bmatrix} a_0 & a_1 & \cdots & a_p & 0 & \cdots & 0 \end{bmatrix}^T
        = \sigma_W^2 \begin{bmatrix} 1 & 0 & \cdots & 0 \end{bmatrix}^T \ \ 
    \end{align}
    That is, the first column of $\Cn{n}^{-1}$---$\bv$ of \eqref{eq:Gohberg_semencul:first_col}---is given by
    $\bv = \sigma_W^{-2} \begin{bmatrix} a_0 & a_1 & \cdots & a_p & 0 & \cdots & 0 \end{bmatrix}^T$.
    Substituting this $\bv$ in \eqref{eq:Gohberg-Semencul} concludes the proof.
\end{IEEEproof}

Moreover, averaging the diagonal elements 
of $\Cn{n}^{-1}$ in \thmref{thm:AR:inverse_covariance} yields the following explicit expression for the normalized Tin~$M_n$.
\begin{corol}[Tin of AR models]
\label{corol:AR:trace-inverse}
    For an AR process of order $p \leq n$, $M_n$, the normalized Tin (\defnref{def:trace-inverse}), is equal to
    \begin{align}
        M_n =  \frac{1}{\sigma_W^2} \sum_{\ell=0}^n \left( 1 - \frac{2\ell}{n} \right) a_\ell^2, 
    \end{align}
    where, by convention, $a_\ell = 0$ for $\ell > p$.
\end{corol}

We now have all the necessary tools for the second proof of \thmref{thm:trace-inverse:monotone} (for non-singular $\Cn{n}$ and $\Cn{n+1}$).

\begin{IEEEproof}[Proof of \thmref{thm:trace-inverse:monotone}]
    Assume that $\Cn{n}$ and $\Cn{n+1}$ are invertible (for the case of singular $\Cn{n+1}$ or $\Cn{n}$ see the proof in \secref{s:proofs}).
    Let $\Cn{n+1}$ be some autocovariance matrix of $n+1$ consecutive samples of a zero-mean WSS process. 
    Then, one can (uniquely) construct an AR process 
    whose coefficients $\{a_0, \ldots, a_n\}$ are determined by the Yule--Walker equations \eqref{eq:YuleWalker}.
    Then, according to \corolref{corol:AR:trace-inverse}, 
    \begin{align}
        \sigma_W^2 \left( M_{n+1} - M_n \right) &= \sum_{\ell=0}^n \left( 1 - \frac{2\ell}{n+1} \right) a_\ell^2
        - \sum_{\ell=0}^n \left( 1 - \frac{2\ell}{n} \right) a_\ell^2
     \\ &= \frac{2}{n(n+1)} \sum_{\ell = 1}^n \ell a_\ell^2 ,
    \end{align}
    where in the expression for $M_{n+1}$ we substituted $a_{n+1} = 0$ since the order of the AR process is smaller than or equal to~$n$.
    
    Thus, $M_{n+1} \geq M_n$, with equality iff all the AR process coefficients but $a_0$ are zero, \ie, iff $\bX_{[n]}$ is white.
    %
\end{IEEEproof}


\section{MInTin versus MaxEnt Completion}
\label{s:mintin-maxent}

In this section, we introduce a spectrum (equivalently, covariance) MinTin completion criterion and contrast this criterion with Burg's MaxEnt criterion. We first derive the infinite-order MinTin completion given $\Cn{m}$ in \secref{s:mintin-maxent:infinite-order}, followed by a derivation of the finite-order MinTin completion out of all possible AR($m$) processes in \secref{s:mintin-maxent:finite-order};
both the infinite- and finite-order MinTin completions are shown to be different, in general, from Burg's MaxEnt completion.

\subsection{Infinite-order Prediction and Completion}
\label{s:mintin-maxent:infinite-order}

We start by deriving an explicit expression for $M_\infty$ and $\LMMSEtwo$~\eqref{eq:two-sided-LMMSE} in \lemref{lem:M1,Minf}, 
followed by a comparison thereof to $M_1$ and $\LMMSEone$. 
We then use a MinTin criterion to complete the entire covariance function (equivalently, spectrum) given its first $L$ values, 
and compare the resulting MinTin spectrum to Burg's MaxEnt spectrum.

\begin{lem}[Infinite-order normalized Tin]
\label{lem:M1,Minf}
    The infinite-order normalized Tin,
    $M_\infty \triangleq \lim_{n \to \infty} M_n$, 
    is 
    equal to the reciprocal of the TSP LMMSE
    $\LMMSEtwo$ \eqref{eq:two-sided-LMMSE}%
    ---the LMMSE in predicting $X_i$ given the entire past and future,
    $\{ X_\ell | \ell \in \ints \backslash \{i\} \}$---and is given by 
    \begin{subequations}
    \label{eq:M_inf}
    \begin{align}
        M_\infty 
        &= \frac{1}{\LMMSEtwo}
    \label{eq:M_inf:LMMSE}
    \\* &= \int_{-1/2}^{1/2} \frac{df}{S\left( \e^{j 2\pi f} \right)}  \:.
    \label{eq:M_inf:spectrum}
    \end{align}
    \end{subequations}
\end{lem}

\begin{IEEEproof}
    Assume first that $\LMMSEtwo > 0$, or, equivalently, that $1/\LMMSEtwo < \infinity$. Assume further, for simplicity, that $n$ is even.
    To prove \eqref{eq:M_inf:LMMSE}, 
    by \colref{col:Tin-LMMSEs} and the symmetry around the center $i = n/2$ of the diagonal, 
    the normalized Tin may be expressed as 
    \label{eq:Minf:proof}
    \begin{align}
        M_n = \frac{1}{n/2} \sum_{i=1}^{n/2} \frac{1}{\LMMSE(i | \except{i}{n})} .
    \label{eq:Minf:proof}
    \end{align}
    Now, since $1/\LMMSEtwo < \infinity$, the (positive) summands inside the sums in \eqref{eq:Minf:proof} converge to $1/\LMMSEtwo$ as $n$ (equivalently, $k$) goes to infinity, and hence so does $M_n$. A similar argument applies when $n$ is odd.
    
    Formula \eqref{eq:M_inf:spectrum} is known from \cite[Ch.~I.10]{Rozanov:Book1967}, \cite{Kay:TSP:TSP1983,Picinbono:TSP:InFrench}, 
    and follows from 
    the facts that the Tin of a matrix is the sum of reciprocals of its eigenvalues
    and 
        that the 
        spectrum of $\Cn{n}$, $\{\lambda^\downarrow_i(\Cn{n})| i \in [n]\}$, converges to the power spectral density $S\left( \e^{j 2\pi f} \right)$ of the process 
        \cite[Ch.~10]{GrenanderSzego:Toeplitz:Book} (see also \cite{Kay:TSP:TSP1983} for a different proof via AR modeling). 
    
    Assume now that $\LMMSEtwo = 0$. Then, by \eqref{eq:Minf:proof}, $M_n$ goes to infinity. 
    On the other hand, there is at least one eigenvalue in the spectrum of $\Cn{n}$ that tends to zero, or equivalently, an eigenvalue of $\Cn{n}^{-1}$ that goes to infinity, and hence the integral in \eqref{eq:M_inf:spectrum} equals infinity. 
\end{IEEEproof}

\begin{remark}
    $M_\infty$ equals the reciprocal of the harmonic mean of the spectrum  $S$ by 
    \eqref{eq:M_inf:spectrum}, 
    while $M_1 = 1/c_0$ equals the reciprocal of the arithmetic mean of the latter since 
    $c_0 = \int_{-1/2}^{1/2} S \left( \e^{j 2\pi f} \right) df$.
    Thus, the inequality $M_1 \leq M_\infty$ between the extremes of \thmref{thm:trace-inverse:monotone} follows immediately from the arithmetic mean--harmonic mean inequality.
\end{remark}

The OSP LMMSE \LMMSEone---the LMMSE in predicting the present given the entire past---equals the geometric mean of the spectrum by the first Szeg\"o limit theorem \cite{Szego:Toeplitz:WSS_prediction_error:1915}, \cite[Ch.~10.8]{GrenanderSzego:Toeplitz:Book}:
\begin{align}
\label{eq:prediction:infinite:1sided}
    \LMMSEone 
    &= \exp \left\{ \int_{-1/2}^{1/2} \log S \left( \e^{j 2\pi f} \right) \right\} .
\end{align}

By \lemref{lem:M1,Minf}, the TSP LMMSE \LMMSEtwo---the LMMSE in predicting the present given the entire past and future---equals the harmonic-mean of the spectrum \eqref{eq:M_inf:spectrum}. 

Comparing the two, the TSP LMMSE is evidently less than or equal to the OSP LMMSE, by the geometric mean--harmonic mean inequality, and the two are equal iff the spectrum $S$ is constant, \ie, iff the process $\{X_n\}$ is white, meaning that the past and the future cannot help in (linearly) predicting the present. Or, using the prediction-gain terms, and denoting the limit of $D_n^G$  \eqref{eq:log-det-ratio}, \eqref{eq:one-sided-prediction} as $n\rightarrow\infty$ by $\overleftarrow{D^G_\infty}$, we have:
\begin{align}
\label{eq:prediction-gains}
\begin{aligned} 
    \overleftarrow{D^G_\infty} &= \half \log \frac{\var{X_1}}{\LMMSEone}
    = \half \int_{-1/2}^{1/2} \log \frac{c_0}{S \left( \e^{j 2\pi f} \right)} df 
 \\ & \leq \half \log \int_{-1/2}^{1/2} \frac{c_0}{ S \left( \e^{j 2\pi f} \right)}  df 
    = \half \log \frac{\var{X_1}}{\LMMSEtwo}  \triangleq \overleftrightarrow{D^G_\infty} , 
\end{aligned}
\end{align}
with equality iff the process is white, \ie, $S \left( \e^{j 2\pi f} \right) = c_0 \;\;\forall f$, in which case $\overleftarrow{D^G_\infty} = \overleftrightarrow{D^G_\infty} = 0$,
where $\overleftarrow{D^G_\infty}$ and $\overleftrightarrow{D^G_\infty}$ are the OSP and TSP gains, respectively, 

\begin{thm}[Infinite-order MinTin completion]
\label{thm:WorstSpectrum}
    Given a PD matrix $\Cn{m}$ for $m \in \nats$ (equivalently, given admissible
    $\{c_0, c_1, \ldots c_{m-1}\}$),
    the minimal $M_\infty$ that is consistent with $\Cn{m}$
    is attained by a zero-mean WSS process with power spectral density of the form 
    \begin{subequations}
    \label{eq:worstSpectrum}
    \begin{align}
        S \left( \e^{j 2\pi f} \right) 
        &= \frac{1}{\sqrt{\sum_{\ell=0}^{m-1} \lambda_\ell \cos(2 \pi \ell f)}} 
    \label{eq:worstSpectrum:cosines}
     \\ &= \frac{\gamma}{\prod_{k=1}^{m-1}|1-\xi_k\e^{-j2\pi f}|}
        \,,
    \label{eq:worstSpectrum:exponents}
    \end{align}
    \end{subequations}
    for the set $\left\{ \lambda_\ell \in \reals | \ell+1 \in [m] \right\}$ (alternatively, $\gamma\in\reals_+$ and the set $\left\{\xi_k\in\compls|k\in[m-1]\right\}$)
    for which 
    the constraints
    \begin{align}
    \label{eq:worstSpectrum:constraints}
        \!\!\!\!
        \int_{-1/2}^{1/2} S \left( \e^{j 2\pi f} \right) \cos(2\pi \ell f) df 
        &= c_\ell, 
        & \ell+1 \in [m] , \quad
    \end{align}
    are satisfied.
    
\end{thm}
\begin{remark}
    If $\Cn{m}$ is singular (PSD but not PD), then $M_m=\infty$, so by \thmref{thm:trace-inverse:monotone} the minimal $M_\infty$ is infinite as well.
\end{remark}

\begin{IEEEproof}
    Due to the stationarity of the process, the given covariance matrix $\Cn{m}$ of $m$ consecutive samples of the process is equivalent to 
    the constraints $\E{X_{i+\ell} X_i} = c_\ell$ for $\ell=0,1,\ldots m-1$.
    By applying the inverse Fourier transform, these constraints are equivalent, in turn, to 
    \eqref{eq:worstSpectrum:constraints}.
    Now, by solving the constrained calculus of variations problem (with variable end points) \cite[Chs.~6 and 12]{GelfandFomin:CalculusOfVariations:Book}, \cite[Ch.~4]{Weinstock:CalculusOfVariations:Book}
    of minimizing \eqref{eq:M_inf:spectrum} under the $m$ integral constraints \eqref{eq:worstSpectrum:constraints}, 
    one arrives at \eqref{eq:worstSpectrum:cosines}; the details may be found in 
    \appref{app:MinTin:WorstSpectrum}. The alternative representation \eqref{eq:worstSpectrum:exponents} easily follows by direct substitution. 
\end{IEEEproof}

    We note that 
    the spectrum that achieves the infinite-order MinTin \eqref{eq:worstSpectrum}
    under the autocorrelation constraints \eqref{eq:worstSpectrum:constraints}
    is different, in general, from Burg's AR($m-1$) MaxEnt solution under the same constraints. According to \lemref{lem:M1,Minf}, the former minimizes the MMSE in estimating the present given the entire past and future (TSP), whereas the latter minimizes the MMSE of the present given only the entire past (OSP). 
    The MaxEnt and MinTin solutions may be further thought of as the closest to a white process in terms of
    the OSP gain $\overleftarrow{D_\infty^G}$ and the TSP gain $\overleftrightarrow{D_\infty^G}$ \eqref{eq:prediction-gains}, respectively,
    given the constraints.
    
    We term a WSS process with spectrum $S \left( \e^{j 2\pi f} \right)$ of the shape \eqref{eq:worstSpectrum} a ``Root-AR" (RAR) process of order $m-1$ [denoted RAR($m-1$)]. Note that, interestingly, any AR($p$) process can be regarded as a particular case of a RAR($2p$) process: Consider an AR($p$) process and let $\sigma_W^2$ and $\{\bar{\xi}_k|k\in[p]\}$ denote the variance of its excitation noise and the $p$ poles of its associated polynomial $A(z)$, respectively. Now consider a set of $2p$ poles $\{\xi_k|k\in[2p]\}$, such that $\xi_k=\xi_{p+k}=\bar{\xi}_k$ for all $k\in[p]$. By using these $2p$ poles with $\gamma=\sigma_W^2$ (and $m=2p+1$) in \eqref{eq:worstSpectrum:exponents}, we obtain a spectrum of a RAR($2p$) process that equals the spectrum of the original AR($p$) process. Note that, as a consequence, given $2p+1$ covariance values $c_0,c_1,\ldots c_{2p}$ that happen to correspond to an AR($p$) process, this AR($p$) process would actually be a RAR($2p$) process satisfying the constraints in \eqref{eq:worstSpectrum:constraints}, and would therefore be the ``MinTin-optimal" process given these $2p+1$ covariance values, which (being an AR process) also happens to coincide with the ``MaxEnt-optimal" process in this case. Naturally, the same holds if the order of this AR process is smaller than $p$. This intriguing property is also reflected in \thmref{thm:mintin-AR} in the sequel.

\begin{remark}
    If nonconsecutive values of the covariance function, 
    $\left\{ c_\ell \middle| \ell+1 \in \cS \subset [m] \right\}$,
    are given in \thmref{thm:WorstSpectrum}, only the corresponding 
    $\left\{ \lambda_{\ell} \middle| \ell+1 \in \cS \subset [m] \right\}$ may be different from zero with the rest equalling zero. 
    This parallels the results for Burg's maximum entropy \cite{Burg:PhD1975:MaximumEntropy} (or equivalently the maximum MMSE given the entire past) with nonconsecutive constraints \cite{RozarioPapoulis:Nonconsecutive-MaximumEntropy:IT1987}.
\end{remark}

We conclude this section by proving that the infinite-order {\em maximum} Tin (MaxTin) completion is always infinite.

\begin{thm}[Infinite-order MaxTin completion]
\label{thm:BestSpectrum}
    Given a PSD covariance matrix $\Cn{m}$ for $m \in \nats$ (equivalently, given admissible $\{c_0, c_1, \ldots c_{m-1}\}$),
    the maximal $M_\infty$ that is consistent with $\Cn{m}$ is infinite.
\end{thm}

Note first that If $\Cn{m}$ is singular, then $M_m=\infty$, so by \thmref{thm:trace-inverse:monotone} the maximal $M_\infty$ is infinite as well.
To prove \thmref{thm:BestSpectrum} for a nonsingular $\Cn{m}$, we essentially need to show that the given covariance sequence can be continued (with an admissible continuation) in a way that leads to a singular covariance matrix $\Cn{n}$ for some $n>m$. To this end, we first show that given any invertible $\Cn{m}$ there is a zero-mean moving-average (MA) process that is consistent with it.

A zero-mean MA process $\{X_i\}$ is a process that 
can be represented as  
\begin{align}
\label{eq:MA:model}
    X_i = \sum_{\ell=0}^q b_\ell W_{i-\ell} , 
\end{align}
where the ``excitation noise" $\{W_i\}$ is a sequence of uncorrelated random variables with zero mean and variance $\sigma_W^2$, $q \in \nats$ is the order of the MA process assuming $b_q \neq 0$, $b_0 = 1$, and $\{b_\ell | b_\ell \in \reals, \ell \in [q] \}$ are the process coefficients. Such a process is denoted by MA($q$).

\begin{thm}[MA modeling with a given covariance]
\label{thm:MA_given_covariance}
    Given a PD covariance matrix $\Cn{m}$ for $m \in \nats$ (equivalently, given corresponding admissible $\{c_0, c_1, \ldots c_{m-1}\}$), there exists an MA process \eqref{eq:MA:model} 
    that is consistent with~it.
\end{thm}

We provide a constructive proof of 
\thmref{thm:MA_given_covariance}
in \appref{app:MA_given_covariance}.

To prove \thmref{thm:BestSpectrum}, 
we construct a matching MA($q$) process that is consistent with $\Cn{m}$ and is guaranteed to exist by \thmref{thm:MA_given_covariance}.
Then, by replacing the excitation noise $\{W_i\}$ of this MA process with a periodic excitation noise with a long enough period (at least $m+q$), such that the random variables within each period are uncorrelated, the result follows. Alternatively, the desired result can be achieved by passing the matching MA($q$) process through a linear time-invariant (LTI) filter with impulse response $h_\ell=\frac{1}{\sqrt{2}} \left( \delta_\ell + \delta_{\ell-m-q} \right)$ with $\delta_\ell$ denoting the Kronecker delta function. The detailed proof of \thmref{thm:BestSpectrum} is available in \appref{app:BestSpectrum}.

\subsection{Finite-order Prediction and Completion}
\label{s:mintin-maxent:finite-order}

Given $\Cn{p+1}$, Burg's MaxEnt principle in estimating the spectrum [equivalently, maximizing \eqref{eq:prediction:infinite:1sided}], yields a zero-mean Gaussian AR process \eqref{eq:AR:model} of order up to $p$, \ie, a spectrum of the form 
\begin{align}
\label{eq:worstSpectrum:MaxEnt}
    S \left( \e^{j 2\pi f} \right) = \frac{\sigma_W^2}{\sum_{k=0}^p \lambda_k \cos(2 \pi k f)} \,,
\end{align}
with $\lambda_0 = \sum_{\ell = 0}^p a_\ell^2$ and $\lambda_k = 2 \sum_{\ell = 0}^{p-k} a_\ell a_{\ell+k}$ for \mbox{$k \in [p]$}, 
where the coefficients 
$\{a_1,\ldots a_p\}$ 
and $\sigma_W^2$ (recall that \mbox{$a_0=1$}) are uniquely determined by the Yule--Walker equations \eqref{eq:YuleWalker}. 

The resulting OSP LMMSE \eqref{eq:prediction:infinite:1sided} reduces, for AR models,~to 
\begin{align}
    \LMMSEone = \sigma_W^2 .
\end{align}
By recalling that an AR($p$) process is a Markov process of order $p$,\footnote{For a Markov process of order $p$, the immediate $p$ past steps and $p$ future steps constitute a sufficient statistic for the present.} and comparing the expressions for the diagonal entries of $\Cn{n}^{-1}$ in \lemref{lem:trace-LMMSEs} and \thmref{thm:AR:inverse_covariance}, we arrive at
\begin{align}
    \LMMSEtwo = \frac{\sigma_W^2}{\sum_{\ell = 0}^p a_\ell^2} \:,
\end{align}
which, as expected, is less than or equal to $\LMMSEone$, since $a_0 = 1$ (see also \cite{Kay:TSP:TSP1983,HsueYagle:TSPvsOSP:TSP1995}).

As implied by \thmref{thm:WorstSpectrum}, the MinTin (RAR) spectrum \eqref{eq:worstSpectrum} is shaped like the square root of a MaxEnt (AR) spectrum \eqref{eq:worstSpectrum:MaxEnt} but with different $\{\lambda_k\}$ values, and is hence different from the MaxEnt spectrum solution \eqref{eq:worstSpectrum}, in general. Somewhat surprisingly, Burg's AR MaxEnt solution is not even the MinTin solution within the class of AR processes of order $p+1$ (or smaller). 

More specifically, assume that a PD covariance matrix $\Cn{p+1}$ (equivalently, $\{c_0,c_1,\ldots c_p\}$) is given,
matching [via the Yule--Walker equations \eqref{eq:YuleWalker}] an AR($p$) process with coefficients $a_0=1, a_1, \ldots, a_p$, with $a_p\ne 0$ (we shall address the possible but less likely case where $\Cn{p+1}$ can be matched by an AR process of a lower order, namely with $a_p=0$, later on). Then, among all AR processes of orders smaller than or equal to $p+1$, consistent with the specified covariance values:
\begin{itemize}
    \item The process maximizing the OSP LMMSE is well-known (by Burg's MaxEnt solution) to be the same AR($p$) process, which implies that its next covariance value is given by $c_{p+1}^{\text{MaxEnt}}\triangleq-\sum_{\ell=1}^p a_\ell c_{p+1-\ell}$.
    \item However, the process maximizing the TSP LMMSE is an AR($p+1$) process, whose next covariance value is specified in the following Theorem.
\end{itemize} 

\begin{thm}[Worst TSP MMSE of finite-order AR models]
\label{thm:mintin-AR}
   Assume 
   a PD covariance matrix $\Cn{p+1}$ for $p-1 \in \nats$ (equivalently, given corresponding admissible $\{c_0, c_1, \ldots c_p\}$)
   and denote the parameters of the implied AR($p$) process $\sigma_W^2$ and $a_0=1, a_1, \ldots, a_p$, assuming $a_p \ne 0$. Then, among all AR processes of orders up to $p+1$, consistent with the specified covariance values, the process maximizing the TSP LMMSE is a zero-mean WSS AR($p+1$) process, whose next covariance value is given by
   \begin{align}
       c_{p+1}^\MaxTSP \triangleq c_{p+1}^{\rm{MaxEnt}}+\left(\alpha-\sign(\alpha)\sqrt{\alpha^2-1}\right)\sigma_W^2,
   \label{eq:mintin:finite-order:cov-seq}
   \end{align}
   where
   \begin{align}
       c_{p+1}^{\rm{MaxEnt}}\triangleq-\sum_{\ell=1}^p a_\ell c_{p+1-\ell}\;\;\;\text{and}\;\;\;
       \alpha\triangleq\frac{\sum_{\ell=0}^p a_\ell^2}{\sum_{\ell=1}^p a_\ell a_{p+1-\ell}}.
   \label{eq:maxent:cov-seq}
   \end{align}
\end{thm}

The proof of \thmref{thm:mintin-AR} is available in \appref{app:mintin-AR}.

\begin{remark}
    Eqs.\ \eqref{eq:mintin:finite-order:cov-seq} and \eqref{eq:maxent:cov-seq} of \thmref{thm:mintin-AR} are still applicable in case the sequence $c_0, c_1, \ldots c_p$ happens to match an AR($p'$) process with $p'<p$, namely when $a_p$, and possibly also $a_{p-1},a_{p-2},\ldots a_{p'+1}$, vanish. In that case, the MaxEnt spectrum (and therefore also the MaxEnt completion) still corresponds to the same AR($p'$) process. However, an interesting distinction applies to the MinTin completion in such a case: If $p'>p/2$, then the MinTin completion still corresponds to an AR($p+1$) process. However, if $p'\le p/2$, then the denominator in \eqref{eq:maxent:cov-seq} vanishes, $\alpha$ becomes infinite, and the correction term in \eqref{eq:mintin:finite-order:cov-seq} vanishes, so that the MinTin completion $c_{p+1}^\MaxTSP$ coincides with the MaxEnt completion $c_{p+1}^{\rm{MaxEnt}}$, and both correspond to an AR($p'$) process. This is in nice agreement with the fact (discussed in Subsection \ref{s:mintin-maxent:infinite-order} above) that the implied RAR($p$) process corresponds to an AR($p'$) process iff $p\ge 2p'$.
\end{remark}

An interesting observation is that the same covariance value $c_{p+1}^\MaxTSP$ also minimizes the Tin of the augmented covariance matrix $\Cn{p+2}$ given $\Cn{p+1}$, as stated in the following theorem, whose proof is available in \appref{app:mintin-C}.

\begin{thm}[Finite-order AR-model MinTin completion]
\label{thm:mintin-C}
    Assume a PD covariance matrix $\Cn{p+1}$ for $p-1 \in \nats$ (equivalently, assume corresponding admissible $\{c_0, c_1, \ldots c_p\}$)
    and denote the parameters of the implied AR($p$) process $\sigma_W^2$ and $a_0=1, a_1, \ldots, a_p$, assuming $a_p \ne 0$.
    Denote by $c^\MinTin_{p+1}$ the value $c_{p+1}$
    that minimizes the Tin of the implied $\Cn{p+2}$. 
    Then, $c_{p+1}^\MinTin = c_{p+1}^\MaxTSP$, where $c_{p+1}^\MaxTSP$ was specified in \thmref{thm:mintin-AR}.
\end{thm}
Note that this property, although rather appealing, is somewhat surprising, since the TSP LMMSE of an AR($p+1$) process maximized in \thmref{thm:mintin-AR} is not directly related to any of the (reciprocal) LMMSEs along the diagonal of $\Cn{p+2}^{-1}$, and in fact, the property stated in \thmref{thm:mintin-C} does not necessarily involve an AR($p+1$) process.

Given $\Cn{p}$ of a WSS process, the expression provided by Ths.~\ref{thm:mintin-AR} and \ref{thm:mintin-C} only identifies the single next covariance value $c_{p}^\MinTin$ (or $c_p^\MaxTSP$) that minimizes the Tin of the implied $\Cn{p+1}$ [or maximizes the TSP LMMSE of the implied AR($p$) process]. Unfortunately, however, it falls short of providing the full (infinite) completion of the covariance sequence for the process that minimizes $M_\infty$ (equivalently, maximizes the TSP LMMSE). \thmref{thm:WorstSpectrum} provides the spectrum of this process, but does not admit a closed-form solution of the associated parameters $\{\lambda_\ell|\ell+1\in[p+1]\}$, let alone the resulting covariance sequence.

A possible approach for trying to approximate the full (infinite) covariance sequence is to compute the subsequent covariance values recursively (using the expression from \thmref{thm:mintin-AR}), advancing one step at a time. However, this ``greedy" approach will not retrieve the exact full covariance sequence: For example, given the first $p+1$ covariance values ($c_0, \ldots, c_{p}$), the first of the two additional covariance values defining the AR($p+2$) process that maximizes the TSP LMMSE would generally be different from $c_{p+1}^\MaxTSP$ of \eqref{eq:mintin:finite-order:cov-seq}.

We illustrate the difference between the TSP LMMSE of the MinTin solution of \thmref{thm:WorstSpectrum}, the recursive (``greedy") application of \thmref{thm:mintin-AR}, and Burg's MaxEnt solution 
in the following example.

\begin{example}
    Since the parameters of the MinTin spectrum (prescribed by \thmref{thm:WorstSpectrum}) do not admit a closed-form solution, we constructed an 
    optimal-MinTin spectrum of order $3$ [RAR($3$), see \eqref{eq:worstSpectrum:exponents}] with 
    predefined 
    coefficients,
    \begin{align}
    \label{eq:Sx_example}
        S \left( \e^{j 2\pi f} \right) = \frac{\gamma}{|1-\xi_1\e^{-j2\pi f}||1-\xi^*_1\e^{-j2\pi f}||1-\xi_2\e^{-j2\pi f}|}\;.\quad\, 
    \end{align}
    We chose $\xi_1=0.97\e^{j0.4\pi}$, $\xi_2=0.99$, and set $\gamma$ (numerically) so as to ensure unit-power ($\gamma\approx 0.4062$ for our choice of $\xi_1$ and $\xi_2$), leading to $c_0=1$,  $c_1\approx 0.6054$, $c_2\approx 0.1324$ and $c_3\approx 0.0904$, 
    which we obtained from the inverse Discrete-Time Fourier Transform (iDTFT) of $S \left( \e^{j 2\pi f} \right)$ by numerical integration.
    
    \begin{figure}[t]
            \centering
            \includegraphics[trim={3.5cm, 8cm, 4.5cm, 8.9cm}, clip, width=0.96\columnwidth]{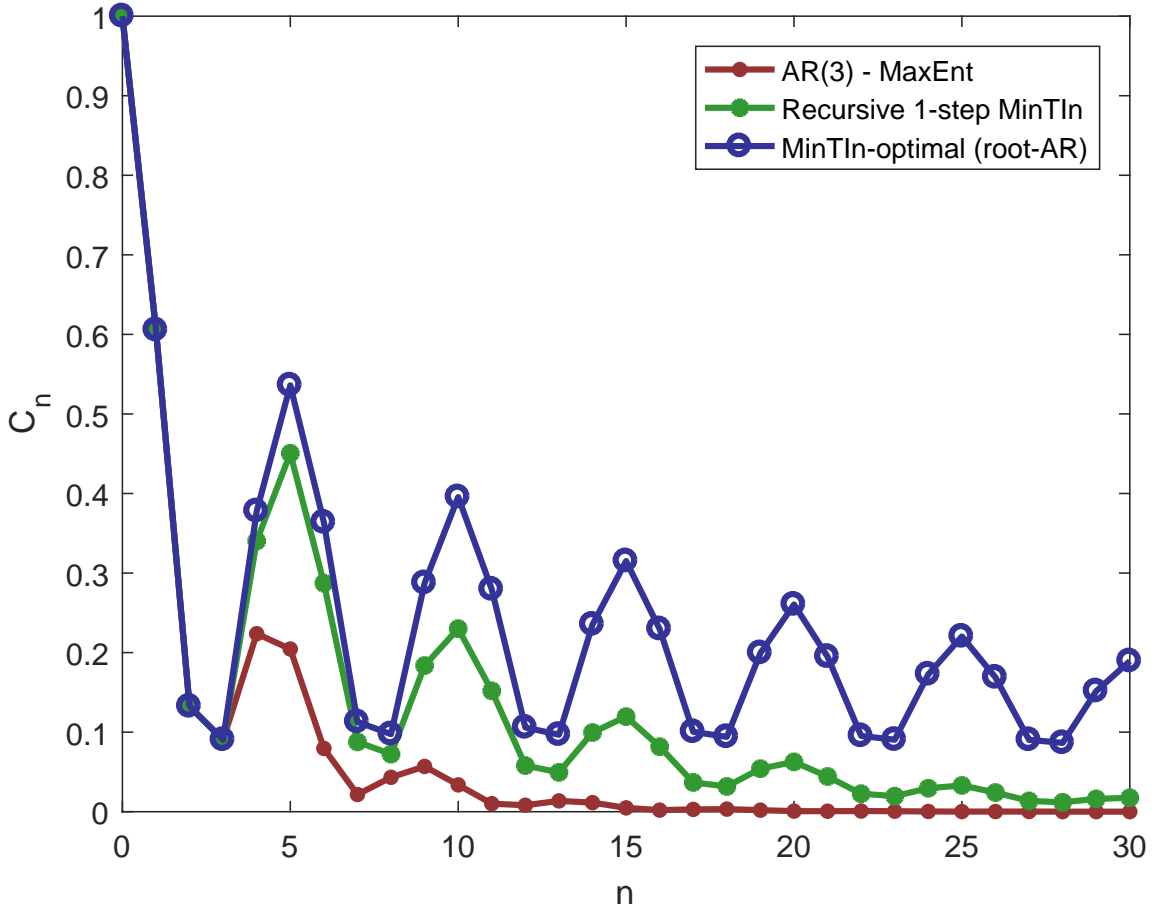}
            \caption{The three covariance sequences completing the four given values of $c_0=1$,  $c_1\approx 0.6054$, $c_2\approx 0.1324$ and $c_3\approx 0.0904$.}
            \label{fig:Cov}
    \end{figure}
    
    \figref{fig:Cov} illustrates the three different completions of this covariance sequences: Burg's MaxEnt completion \eqref{eq:maxent:cov-seq}; the sequence obtained by recursive (greedy) application of \eqref{eq:mintin:finite-order:cov-seq} of \thmref{thm:mintin-AR} for $p=1,2,\ldots$; and the infinite-order MinTin (MaxTSP) sequence of \thmref{thm:WorstSpectrum} obtained from the iDTFT of $S \left( \e^{j 2\pi f} \right)$ of  \eqref{eq:Sx_example}.
    
    \begin{figure}[t]
            \centering
            \includegraphics[trim={3.5cm, 8cm, 4.5cm, 8.9cm}, clip, width=0.96\columnwidth]{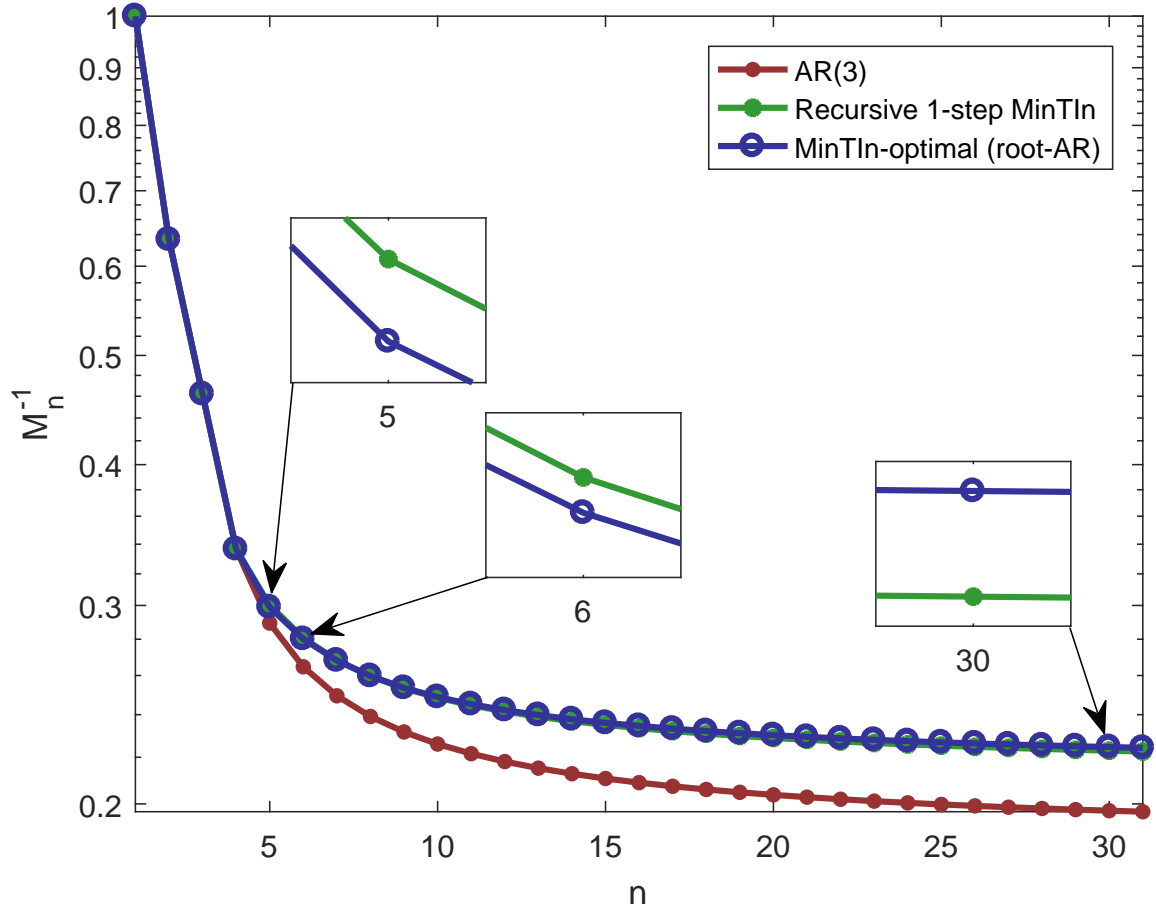}
            \caption{The reciprocal normalized Tin values $M_n^{-1}$ versus $n$, corresponding to the three sequences in \figref{fig:Cov}. The three zoomed-in inserts are positioned around $n=5$, $n=6$ and $n=30$, and all have the same vertical span of $\sim 0.0026$.}
            \label{fig:Mn}
    \end{figure}
    
    \figref{fig:Mn} shows the respective reciprocal normalized Tin 
    $M_n^{-1}$ 
    (the harmonic mean of the TSP LMMSEs) of $\Cn{n}$ vs.\ $n$ for the covariance sequences depicted in \figref{fig:Cov}. As expected, for $n=5$ the recursive (``greedy'') MinTin completion of \eqref{eq:mintin:finite-order:cov-seq} attains the largest possible $M_5^{-1}$ (largest possible harmonic mean of the five TSP LMMSEs in $\Cn{5}$), as guaranteed by \thmref{thm:mintin-C}. Even for $n=6$, the recursive completion still ``outperforms" the infinite-order MinTin [RAR($3$)] completion, since the latter is only asymptotically ``optimal'' (in terms of minimum Tin)---as indeed observed from the plots at the higher values of $n$. Note, though, that the differences between the values of $M_n^{-1}$ between the recursive completion sequence and the infinite-order MinTin covariance sequence are very small, and are only observed in the zoomed-in inserts in \figref{fig:Mn}.
    
    \begin{figure}[t]
            \centering
            \includegraphics[trim={3.5cm, 8cm, 4.5cm, 8.9cm}, clip, width=0.96\columnwidth]{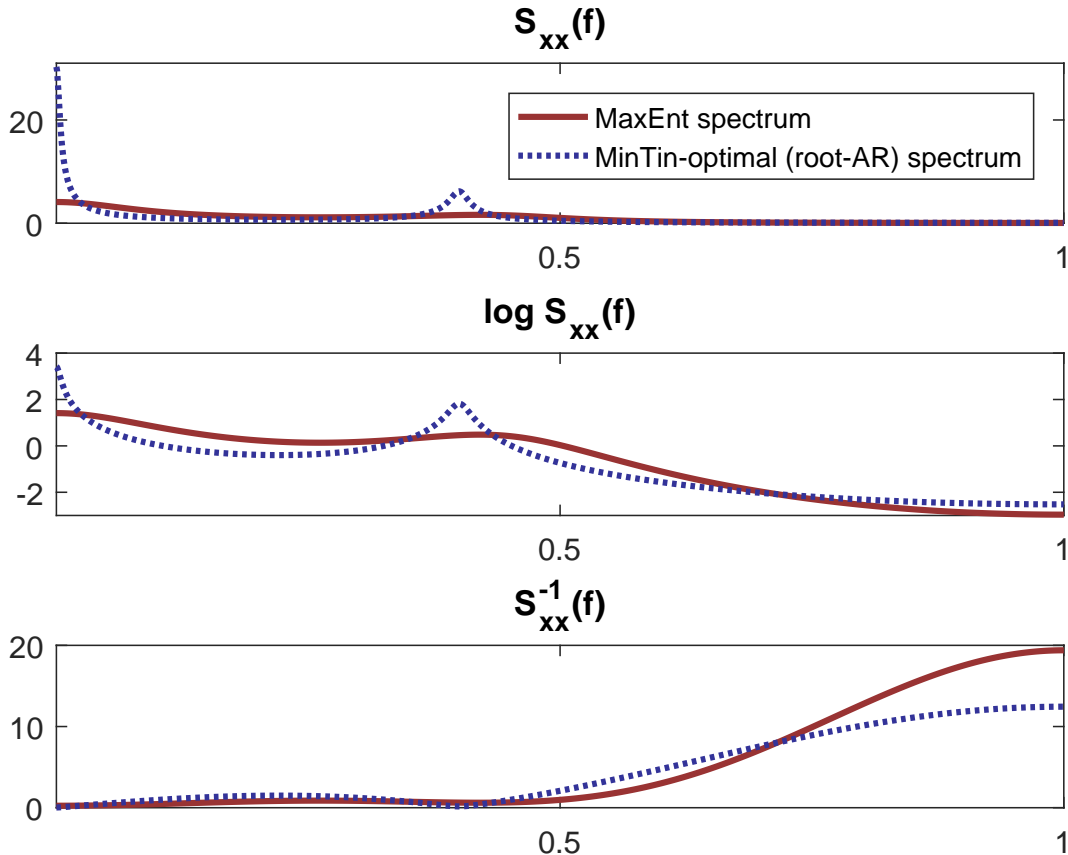}
            \caption{The maxEnt and MinTin spectra in linear, log and inverse scales.}
            \label{fig:Spectra}
    \end{figure}
    
    \figref{fig:Spectra} illustrates the differences between the MaxEnt (Burg's) spectrum and the optimal MinTin (RAR) spectrum (both subject to the first $4$ covariance values 
    specified above), on three different scales: Linear, Logarithmic and Inverse. Interestingly, while the MaxEnt spectrum appears more ``flat" (``white") than the MinTin spectrum on the Linear and Logarithmic scales, the converse is true for the inverse scale.
\end{example}


\section{Trace--Inverse for Non-Stationary Processes}
\label{s:non-stationary}

In this section, we part with the stationarity assumption and assume 
a general zero-mean process $\{X_i\}$ (not necessarily WSS).
We 
derive a monotonicity result for non-stationary processes in the spirit of \thmref{thm:trace-inverse:monotone} (and akin to those of \cite[Sec.~II-B]{DemboCoverThomas:inequalities:IT91}, \cite[Ch.~17.6]{CoverBook2Edition} for average entropies and average conditional entropy rates). 

\begin{defn}
\label{def:M_k^n}
    The normalized $k$-out-of-$n$ Tin, $M_k^{(n)}$, is defined as the per-symbol average of the Tins of the autocovariance matrices of all possible $k$-element subsets of $\bX_{[n]}$:
    \begin{align}
        M_k^{(n)} \triangleq \frac{1}{\binom{n}{k}} \sum_{\cS \subseteq [n] : |\cS| = k} \frac{1}{k} \trace{\Cn{S}^{-1}} .
    \end{align}
\end{defn}

\begin{thm}[Monotonicity of normalized-Tin of subsets] 
\label{thm:non-WSS:Mn:monotonicity}
    The sequence $\left\{ M_k^{(n)} \middle| k \in [n] \right\}$ is monotonically 
    non-decreasing in $k \in [n]$ for a fixed $n \in \nats$. Namely, 
    \begin{align}
    \label{eq:non-stationary:trase-inverse:monotonoe}
        M_k^{(n)} &\leq M_{k+1}^{(n)}, &\forall n &\in \nats, & \forall k \in [n-1].
    \end{align}
\end{thm}
\begin{remark}
    The inequality \eqref{eq:non-stationary:trase-inverse:monotonoe} does not specialize to \eqref{eq:trace-inverse:monotone} even when the process $\{X_k\}$ is WSS. 
\end{remark}

\begin{IEEEproof}
    The following set of inequalities proves \thmref{thm:non-WSS:Mn:monotonicity}.
    \begin{subequations}
    \label{eq:proof:M_k^n}
    \begin{align}
        & M_{k+1}^{(n)} \triangleq \frac{1}{\binom{n}{k+1}} \sum_{\cS \subseteq [n] : |\cS| = k+1} \frac{1}{k+1} \trace{\Cn{S}^{-1}}
    \label{eq:proof:M_k^n:def}
     \\ &= \frac{1}{\binom{n}{k+1}} \sum_{\cS \subseteq [n] : |\cS| = k+1} \frac{1}{k+1} \sum_{i \in \cS} \frac{1}{\LMMSE(i | \cS \backslash \{i\})}
    \label{eq:proof:M_k^n:trace-LMMSE_n+1}
     \\ &\geq \frac{1}{\binom{n}{k+1}} \sum_{\cS \subseteq [n] : |\cS| = k+1} \frac{1}{k+1} \sum_{i \in \cS}
      \frac{1}{k} \sum_{\ell \in \cS \backslash \{i\}} \frac{1}{\LMMSE(i | \cS \backslash \{i,\ell\})} \qquad 
    \label{eq:proof:M_k^n:measurment_decrease_LMMSE}
     \\ &= \frac{1}{\binom{n}{k+1}} \frac{1}{k+1} \sum_{\cS \subseteq [n] : |\cS| = k} \frac{1}{k} \sum_{i \in \cS} (n - k) \frac{1}{\LMMSE(i | \cS \backslash \{i\})}
    \label{eq:proof:M_k^n:rearrange}
     \\ &= \frac{1}{\binom{n}{k}} \sum_{\cS \subseteq [n] : |\cS| = k} M_k^{(n)} ,
    \label{eq:proof:M_k^n:trace-LMMSE_n}
    \end{align}
    \end{subequations}
    where \eqref{eq:proof:M_k^n:def} is by \defnref{def:M_k^n}, 
    \eqref{eq:proof:M_k^n:trace-LMMSE_n+1} and \eqref{eq:proof:M_k^n:trace-LMMSE_n} follow from \lemref{lem:trace-LMMSEs}, 
    \eqref{eq:proof:M_k^n:measurment_decrease_LMMSE} follows from \lemref{lem:LMMSE_more_measurments}, 
    \eqref{eq:proof:M_k^n:rearrange} follows from rearranging the terms and noting that 
    every reciprocal-LMMSE term in the double-summation in \eqref{eq:proof:M_k^n:rearrange}
    appears $(n-k)$ times 
    in the triple-summation in \eqref{eq:proof:M_k^n:measurment_decrease_LMMSE}.
\end{IEEEproof}


\section{Discussion}
\label{s:summary}

In this work, the normalized Tin was shown to be intimately related to two-sided prediction and monotonically non-decreasing with the order, similarly to the joint (differential) entropy which is monotonically non-increasing.
Furthermore, MinTin serves as an alternative criterion to the widely used MaxEnt criterion (that is associated with one-sided prediction) for spectrum estimation/completion. 
The MinTin criterion may be used as a basis in research domains where the cost of a system with causal constraints is put up against its non-causal counterpart (causal versus non-causal regret), 
e.g., 
sequential universal compression, online least-square estimation and control;
as well as domains where non-causal operation is possible, \eg, B-frames in video coding. See also the discussion in \cite{HaikinZamirGavish:MANOVA:PNAS2017} in the context of performance measures for frames. 

Finally, we note that extending the fully-observable (``pure prediction'') scenario considered in this manuscript
to a partially-observable one is an interesting direction for future research.


\appendices


\section{Proof of equality condition in \thmref{thm:trace-inverse:monotone} via \\ MMSE estimation} 
\label{app:tin-monotonicity:equality-cond}

Assume 
that $\Cn{n}$ and $\Cn{n+1}$ are invertible (the cases of singular $\Cn{n}$ and/or $\Cn{n+1}$ have been treated separately in \secref{s:proofs}).
To attain equality in \eqref{eq:trace-inverse:monotone}, \eqref{eq:LMMSE-mean:LMMSEs} must hold with equality for all $\ell \in [n]$,  
whereas \eqref{eq:traces-inequality:prev-ineqs} suggests further that 
\begin{align}
    \frac{1}{\LMMSE(i | \except{i}{n})} &= \frac{1}{\LMMSE(k | \except{k}{n+1})} = M_n 
\end{align}
for all $i \in [n]$ and $k \in [n+1]$;
by \lemref{lem:trace-LMMSEs}, 
this implies
\begin{align} 
\label{eq:equality-cond:traces}
    \trace{[\Cn{n+1}^{-1}]_{\except{i}{n+1}, \except{i}{n+1}}} &= \trace{\Cn{\except{i}{n+1}}^{-1}},
    \forall i \in [n+1]. 
\end{align}
We shall next prove that \eqref{eq:equality-cond:traces} holds iff the vector $\bX_{[n+1]}$ is white.
To this end, we shall first prove that
\begin{align} 
\label{eq:equality-cond:traces:ineq}
    \trace{[\Cn{n+1}^{-1}]_{[n],[n]}} \geq \trace{\Cn{n}^{-1}} 
\end{align}
with equality
iff $\Cn{n+1,[n]} = \veczero^T$,
\ie, iff $X_{n+1}$ is uncorrelated with $\bX_{[n]}$. 

By \lemref{lem:InverseMatrix:Schur}, 
the upper-left $n \times n$ submatrix of $\Cn{n}^{-1}$ is equal~to
\begin{align}
    [\Cn{n+1}^{-1}]_{[n],[n]} &= \left( \Cn{n} - \Cvec{n+1,[n]}^T C_{n+1,n+1}^{-1} \Cvec{n+1,[n]} \right)^{-1} ,
\nonumber
\end{align}
or, equivalently, 
\begin{align}
    \Cn{n} = \left([\Cn{n+1}^{-1}]_{[n],[n]}\right)^{-1} + C_{n+1,n+1}^{-1} \Cvec{n+1,[n]}^T \Cvec{n+1,[n]}.
\end{align}
Now note that $\Cn{n}$ is PD, $\left([\Cn{n+1}^{-1}]_{[n],[n]}\right)^{-1}$
is PSD being an estimation-error covariance matrix (and hence so is its inverse), 
and $C_{n-1,n-1}^{-1} \CwithOthers{n+1}{n+1}^T \CwithOthers{n+1}{n+1}$ is rank-one PSD.
Hence, 
\begin{align}
    \lambda^\downarrow_i \left( \left([\Cn{n+1}^{-1}]_{[n],[n]}\right)^{-1} \right)
    &\leq \lambda^\downarrow_i \left( \Cn{n} \right), 
    & i \in [n], 
\end{align}
where at least one of the inequalities is strict unless $C_{n+1,n+1}^{-1} \Cvec{n+1,[n]}^T \Cvec{n+1,[n]} = \bzero$ \cite[Corollary~4.3.9]{HornJohnsonBook:MatrixAnalysis}, \ie, 
unless $\Cvec{n+1,[n]} = \veczero^T$.
Summing the reciprocal eigenvalues for the traces of the respective inverses, we obtain
\eqref{eq:equality-cond:traces:ineq} with equality iff $\Cvec{n+1,[n]} \equiv \CwithOthers{n+1}{n+1} = \veczero^T$.
 
By the same arguments and after rearranging the entries of the vector $X_{[n+1]}$, one arrives at 
\begin{align} 
    \trace{[\Cn{n+1}^{-1}]_{\except{i}{n+1},\except{i}{n+1}}} &\leq \trace{\Cn{\except{i}{n+1}}^{-1}}, 
    &i \in [n+1], 
\nonumber 
\end{align}
with equality iff $\CwithOthers{i}{n+1} = \bzero$. 
Thus, for \eqref{eq:equality-cond:traces} to hold, the entries of $\bX_{[n+1]}$ must be uncorrelated.
Recalling that $\{X_i\}$ is a WSS process with variance $c_0$, 
concludes the proof.


\section{Proof of \thmref{thm:AR:inverse_covariance} for $p = n$}
\label{app:inverse-covariance:p=n}

We are left with deriving explicit expressions for the elements of $\Cn{p}^{-1}$ in terms of the process coefficients $\left\{ a_\ell \middle| \ell \in [p] \right\}$ and $\sigma_W^2$. To this end, let us consider $\Cn{p+1}^{-1}$ first. Note that with $n=p+1$, we can readily use the result of \thmref{thm:AR:inverse_covariance} which we have already proved for $p<n$. To simplify the exposition, we also assume, w.l.o.g., the scaling convention $c_0=1$ (any different scaling can then be accounted for in $\sigma_W^2$).

Note further that $\Cn{p+1}$ admits the four-blocks structure
\begin{align}
    \Cn{p+1}=\begin{bmatrix}
    \Cn{p} & \Cvec{[p],p+1} \\
    \Cvec{[p],p+1}^T & 1
    \end{bmatrix},
\end{align}
so, using 
\lemref{lem:InverseMatrix:Schur}, we may write
\begin{equation}
\label{eq:B:four_blocks}
    \Cn{p+1}^{-1}=\begin{bmatrix}
    \bQ^{-1} & -\bQ^{-1}\Cvec{[p],p+1} \\
    -\Cvec{[p],p+1}^T \bQ^{-1} & 1+\Cvec{[p],p+1}^T \bQ^{-1}\Cvec{[p],p+1}
    \end{bmatrix},
\end{equation}
where $\bQ \triangleq \Cn{p+1} \backslash 1 = \Cn{p}-\Cvec{[p],p+1}\Cvec{[p],p+1}^T$,
implying
\begin{align}
    \Cn{p}=\bQ+\Cvec{[p],p+1}\Cvec{[p],p+1}^T,
\end{align}
which in turn implies, by \eqref{eq:matrix-inversion-lemma}
(this special case is also known as the Sherman--Morrison formula; see, e.g., \cite[Ch.~0.7]{HornJohnsonBook:MatrixAnalysis}), 
\begin{align}
    \Cn{p}^{-1}=\bQ^{-1}-\frac{(\bQ^{-1}\Cvec{[p],p+1})(\bQ^{-1}\Cvec{[p],p+1})^T}{1+\Cvec{[p],p+1}^T \bQ^{-1}\Cvec{[p],p+1}}.
\end{align}
This relation can be used to obtain explicit expressions for the elements of $\Cn{p}^{-1}$ based on $\Cn{p+1}^{-1}$ \eqref{eq:B:four_blocks} as follows:
\begin{itemize}
    \item The elements of $\bQ^{-1}$ can be readily read off the upper-left $p\times p$ block of $\Cn{p+1}^{-1}$.
    \item The vector $\bQ^{-1}\Cvec{[p],p+1}$ can be readily read off the upper $p$ elements of the last column of $\Cn{p+1}^{-1}$.
    \item $1+\Cvec{[p],p+1}^T \bQ^{-1}\Cvec{[p],p+1}$ is simply the lower-right element of $\Cn{p+1}^{-1}$.
\end{itemize}
Consequently, we have
\begin{equation}
\label{eq:B:iCp_intermsof_iCpp1}
    \left[\Cn{p}^{-1}\right]_{i,j}=\left[\Cn{p+1}^{-1}\right]_{i,j}-\frac{\left[\Cn{p+1}^{-1}\right]_{i,p+1}\left[\Cn{p+1}^{-1}\right]_{j,p+1}}{\left[\Cn{p+1}^{-1}\right]_{p+1,p+1}}.
\end{equation}
Note now that, according to \thmref{thm:AR:inverse_covariance} (with $p<n$), we have
\begin{align}
    \left[\Cn{p+1}^{-1}\right]_{i,p+1}\sigma_W^2&=\sum_{\ell=0}^{i-1}a_{\ell} a_{\ell+p+1-i}-\sum_{\ell=1}^i a_{\ell} a_{\ell+p+1-i}
 \\ &=a_0 a_{p+1-i}-a_i a_{p+1}=a_{p+1-i}, \;\;\; \forall i\in[p+1]
\end{align}
(since $a_0=1$ and $a_{p+1}=0$), and, in particular,
\begin{align}
    \left[\Cn{p+1}^{-1}\right]_{p+1,p+1}\sigma_W^2=a_0=1.
\end{align}
Substituting into \eqref{eq:B:iCp_intermsof_iCpp1}, we get
\begin{align}
    \left[\Cn{p}^{-1}\right]_{i,j}\sigma_W^2&=\left[\Cn{p+1}^{-1}\right]_{i,j}\sigma_W^2-a_{p+1-i}a_{p+1-j}
 \\ &=\sum_{\ell=0}^{i-1}a_{\ell} a_{\ell+j-i}-\sum_{\ell=p+2-j}^{p+1+i-j}a_{\ell} a_{\ell+j-i}-a_{p+1-i}a_{p+1-j}
\nonumber 
 \\ &=\sum_{\ell=0}^{i-1}a_{\ell} a_{\ell+j-i}-\sum_{\ell=p+1-j}^{p+i-j}a_{\ell} a_{\ell+j-i},
\end{align}
which is exactly the expression of \thmref{thm:AR:inverse_covariance} for $p=n$,
where in the last transition the change in the upper limit in the second sum is due to the fact that $a_\ell=0$ for $\ell>p$, 
and the change in the lower limit accounts for the subtraction of the additional term $a_{p+1-i}a_{p+1-j}$.


\section{Detailed proof of the minimizer of $M_\infty$ in \thmref{thm:WorstSpectrum}}
\label{app:MinTin:WorstSpectrum}

\underline{\textit{Necessity:}}
Assume that the power spectral density $S\left(e^{j 2 \pi f} \right)$ (simply denoted $S$ in here for brevity)
that minimizes 
\eqref{eq:M_inf:spectrum} under the $n$ integral constraints \eqref{eq:worstSpectrum:constraints} is continuously differentiable.\footnote{A minimizing power spectral density exists as it is bounded below by, \eg, $M_1=C_1^{-1}$ by \thmref{thm:trace-inverse:monotone} and a spectrum that satisfies the constraints \eqref{eq:worstSpectrum:constraints} by, \eg, an appropriate autoregressive process (recall \secref{s:AR}) whose coefficients may be determined via the Yule--Walker equations \eqref{eq:YuleWalker}.} 
Then, by \cite[Th.~1, Sec.~12]{GelfandFomin:CalculusOfVariations:Book}, \cite[Ch.~4.2]{Weinstock:CalculusOfVariations:Book}, 
it is an (unconstrained) extremum of the functional 
\begin{align}
\label{eq:calculusOfVariations:J}
    J \left[S\right] \triangleq \int_{-1/2}^{1/2} L \left( f, S \right) df , 
\end{align}
where $L$ is the Lagrangian 
\begin{align} 
\label{eq:Lagrangian}
    L \left( f, S \right) \triangleq \frac{1}{S} + \sum_{k=0}^{n-1} \lambda_k S \cos(2\pi k f), 
\end{align}
and therefore the first variation of $J$ \eqref{eq:calculusOfVariations:J} must equal zero, \ie, $L$ must satisfy the Euler--Lagrange equation
\begin{align}
\label{eq:Euler-Lagrange}
    \frac{\partial L}{\partial S} = \frac{d}{d f} \frac{\partial L}{\partial S'}
\end{align}
for some Lagrange multipliers $\{\lambda_k \in \reals | k+1 \in [n]\}$ that are determined 
from the constraints \eqref{eq:worstSpectrum:constraints} and variable end-point conditions 
\begin{align}
\label{eq:variable-end-points}
    \frac{\partial L}{\partial S'} \Bigg|_{f=\pm 1/2} = 0 ,
\end{align}
where $S'$ denotes the derivative of $S$ with respect to $f$. 
For $L$ of \eqref{eq:Lagrangian}, Eq.~\eqref{eq:Euler-Lagrange} reduces to 
\begin{align}
    -\frac{1}{S^2} + \sum_{k=0}^{n-1} \lambda_k \cos(2\pi k f) = 0,
\end{align}
or, equivalently, to \eqref{eq:worstSpectrum:cosines}, 
and the conditions of \eqref{eq:variable-end-points} are trivially satisfied. 
To satisfy the conditions for 
being
an extremum of the problem, the resulting $S$ 
should not be an extremum of any of the constraints \eqref{eq:worstSpectrum:constraints}, \ie,
it need not satisfy \eqref{eq:Euler-Lagrange} and \eqref{eq:variable-end-points} with respect to $S \left( \e^{j 2\pi f} \right) \cos(2\pi k f)$ in lieu of $L$ for any $k+1 \in [n]$, which is indeed the case.

\underline{\textit{Sufficiency:}}
Since $M_\infty$ is bounded from below, \eg, by $M_1$, but is unbounded from above 
by \thmref{thm:BestSpectrum},
we conclude that the derived extremum is indeed a minimum. 
Alternatively, one may easily verify by direct calculation that the second variation \cite[Ch.~ 24, Th.~2]{GelfandFomin:CalculusOfVariations:Book} of the functional $J$~\eqref{eq:calculusOfVariations:J} is equal to 
\begin{align}
    \delta^2 J[h] = \int_{-1/2}^{1/2} \frac{2 h^2\left(e^{j 2 \pi f} \right)}{S^3\left(e^{j 2 \pi f} \right)} df 
\end{align}
and is strongly positive at the minimizing solution $S$ \eqref{eq:worstSpectrum}.\footnote{$\delta^2 J[h]$ is strongly positive if there exists $\kappa > 0$ such that $\delta^2 J[h] \geq \kappa \norm{h}^2$ for all $h$.} 


\section{Proof of \thmref{thm:MA_given_covariance}}
\label{app:MA_given_covariance}

%
Consider the Toeplitz symmetric matrix $\tCn{m}$, whose first-row elements are given by 
\begin{align}
\label{eq:PrecodedAutocovariance}
    \left[ \tCn{m} \right]_{1,\ell+1} &= \frac{k}{k - \ell} \left[ \Cn{m} \right]_{1,\ell+1} , & \ell+1 \in [m],
\end{align}
where $k\in\nats$ is a large enough constant, $k\ge m$, such that $\tCn{m}$ is PD. Such a value of $k$ must exist: By Gershgorin's Circle theorem \cite[Ch.~6]{HornJohnsonBook:MatrixAnalysis}, which can be used to bound the perturbation of the eigenvalues of a matrix by the total perturbation of its rows, the fact that $\Cn{m}$ is PD implies that $\tCn{m}$ \eqref{eq:PrecodedAutocovariance} is PD as well if the perturbation is small enough, namely, if $k$ is large enough.

Therefore, there exists a zero-mean WSS AR process \eqref{eq:AR:model} that is consistent with $\tCn{m}$
and whose coefficients can be found via the Yule--Walker equations \eqref{eq:YuleWalker}.
Finally, obtain an MA process of order $k-1$ \eqref{eq:MA:model} by multiplying 
the autocovariance function of this AR process by a triangular (Bartlett--Fej\'{e}r) window of support size $2k-1$, 
such that its autocovariance matrix $\Cn{m}^\mathrm{MA}$ is equal to $\Cn{m}$ since 
\begin{align}
    \left[ \Cn{m}^\mathrm{MA} \right]_{1,\ell+1} &= \frac{k - \ell}{k} \left[ \tCn{m} \right]_{1,\ell+1}
 = \left[ \Cn{m} \right]_{1,\ell+1},
     & \forall \ell+1 \in [m],
\nonumber
\end{align}
as desired, where the 
second equality
is due to \eqref{eq:PrecodedAutocovariance}.
Note that the autocovariance function of the resulting MA process is admissible, since its Fourier Transform (spectrum) is the convolution of the spectrum of the AR process with the Fourier transform of the triangular window, and since both are non-negative functions, so is their convolution.


\section{Proof of \thmref{thm:BestSpectrum}}
\label{app:BestSpectrum}

By \thmref{thm:MA_given_covariance}, there exists a matching MA($q$) process with some coefficients 
$\{b_0, b_1,\ldots b_q \}$, such that the covariance of $m$ consecutive samples thereof is $\Cn{m}$. 

Now, by exciting the same MA filter (whose impulse response is $h_\ell=b_\ell$ for $\ell=0,1,\ldots q$ and $h_\ell=0$ otherwise)
by an $(m+q)$-periodic excitation noise $W_i$ (in lieu of the standard white excitation noise), such that $\{W_i | i \in [L+q]\}$ are mutually uncorrelated, with zero mean and variance $\sigma_W^2$, the resulting periodic process satisfies the covariance constraint $\Cn{m}$ but has $M_\infty = \infty$ by periodicity and \lemref{lem:M1,Minf}.

Alternatively, by passing the matching MA($q$) process through an LTI filter with impulse response $h_\ell=\frac{1}{\sqrt{2}} \left( \delta_\ell + \delta_{\ell-m-q} \right)$ with $\delta_\ell$ denoting the Kronecker delta function, one attains a new zero-mean MA process of order $m+q$ 
with coefficients 
\begin{align}
    \tb_\ell = \frac{1}{\sqrt{2}}
    \begin{cases}
        b_\ell, & \ell=0,1,\ldots q; 
     \\ b_{\ell-m-q}, & \ell=m+q,m+q+1,\ldots m+2q; 
     \\ 0, & \mathrm{otherwise}.
    \end{cases}
\end{align}
Clearly, the covariance of $m$ consecutive samples of this process is $\Cn{m}$. 
Furthermore, the PSD of this process is equal to that of the original MA process multiplied by $1 + \cos(2\pi f(m+q))$, and hence, 
by \lemref{lem:M1,Minf}, $M_\infty = \infty$ for this process. 


\section{Proof of \thmref{thm:mintin-AR}}
\label{app:mintin-AR}

    To simplify the exposition, we define the following vectors:
    \begin{align}
        \bc_p  &\triangleq [\Cn{p+1}]_{[p+1],1} = \begin{bmatrix}c_0 & c_1 & \cdots & c_{p}\end{bmatrix}^T, \\
        \bbc_p &\triangleq [\Cn{p+1}]_{\except{1}{p+1},1} = \begin{bmatrix}c_1 & c_2 & \cdots & c_{p}\end{bmatrix}^T, \\
        \bbbc_p &\triangleq [\Cn{p+1}]_{[p],p+1} = \begin{bmatrix}c_p & c_{p-1} & \cdots & c_{1}\end{bmatrix}^T.
    \end{align}
    Likewise, let us define
    \begin{align}
        \ba &\triangleq \begin{bmatrix}1 & a_1 & \cdots & a_{p}\end{bmatrix}^T, \\
        \bba &\triangleq \begin{bmatrix}a_1 & a_2 & \cdots & a_{p}\end{bmatrix}^T, \\
        \bbba &\triangleq \begin{bmatrix}a_p & a_{p-1} & \cdots & a_{1}\end{bmatrix}^T.
    \end{align}
    Due to the Yule--Walker equations \eqref{eq:YuleWalker} we have $\sigma_W^2=\ba^T\bc_p$, so we can write the TSP LMMSE of the AR($p$) process 
    (recall \lemref{lem:M1,Minf}) 
    as
    \begin{align}
    \label{eq:Lp}
        \LMMSEtwoP{p} &\triangleq \frac{\sigma_W^2}{\sum_{\ell=0}^p a_\ell^2}
        = \frac{\ba^T\bc_p}{\ba^T\ba}
        = \frac{c_0+\bba^T\bbc_p}{1+\bba^T\bba}
        = \frac{c_0-\bbc_p^T\Cn{p}^{-1}\bbc_p}{1+\bbc_p^T\Cn{p}^{-2}\bbc_p}, \ \ 
    \end{align}
    where for eliminating $\bba$ in the last transition we substituted the Yule--Walker relation $\bba=-\Cn{p}^{-1}\bbc_p$.
    
    Moving to an AR($p+1$) process, our goal is to choose $c_{p+1}$ so as to maximize its TSP LMMSE $\LMMSEtwoP{p+1}$. To this end, we shall express $c_{p+1}$ as $c_{p+1}=c_{p+1}^{\rm{MaxEnt}}+\mu$ and maximize w.r.t.\ $\mu$. Let us denote
    \begin{align}
        \bbc_{p+1}=\begin{bmatrix}c_1\\\vdots\\c_p\\c_{p+1}\end{bmatrix}
        \triangleq
        \begin{bmatrix}c_1\\\vdots\\c_p\\c_{p+1}^{\rm{MaxEnt}}\end{bmatrix}+
        \begin{bmatrix}0\\\vdots\\0\\\mu\end{bmatrix}\triangleq
        \tbc_{p+1}+\bmu.
    \end{align}
The TSP LMMSE for this AR($p+1$) process is then given by substituting $p$ with $p+1$ in \eqref{eq:Lp}:
\begin{subequations}
\noeqref{eq:Lpp1a:2}
\begin{align}
    \LMMSEtwoP{p+1} &= \frac{c_0-\bbc_{p+1}^T\Cn{p+1}^{-1}\bbc_{p+1}}{1+\bbc_{p+1}^T\Cn{p+1}^{-2}\bbc_{p+1}}
\label{eq:Lpp1a}
 \\ &= \frac{c_0-(\tbc_{p+1}+\bmu)^T\Cn{p+1}^{-1}(\tbc_{p+1}+\bmu)}{1+(\tbc_{p+1}+\bmu)^T\Cn{p+1}^{-2}(\tbc_{p+1}+\bmu)}
\label{eq:Lpp1a:2}
\end{align}
\end{subequations}
(note that $\Cn{p+1}$ is still fully available from the given covariance values up to $c_p$). The key observation for proceeding now, is that since $\tbc_{p+1}$ continues the original AR($p$) process, we have
\begin{align}
    \Cn{p+1}^{-1}\tbc_{p+1}=-\begin{bmatrix}\bba\\0\end{bmatrix}
\end{align}
(where $\bba$ still relates to the original AR($p$) process), so exploiting the mostly-zeros structure of $\bmu$, the LMMSE expression simplifies into
\begin{align}
\label{eq:Lpp1}
    \!\!\!\!
    \LMMSEtwoP{p+1}=\frac{c_0+\bba^T\bbc_p-\mu^2\left[\Cn{p+1}^{-1}\right]_{p+1,p+1}}
    {1+\bba^T\bba+2\mu\left[\Cn{p+1}^{-2}\tbc_{p+1}\right]_{p+1}+\mu^2\left[\Cn{p+1}^{-2}\right]_{p+1,p+1}}.\quad
\end{align}
To obtain the elements involving the inversion of $\Cn{p+1}$, we first partition the (symmetric) inverse as
\begin{align}
    \Cn{p+1}^{-1}=\begin{bmatrix}\bQ & \bq\\ \bq^T & q\end{bmatrix},
\end{align}
where $\bQ$ is some $(p-1)\times(p-1)$ matrix, $\bq$ is some $(p-1)\times 1$ vector and $q$ is some scalar. Using the Yule--Walker equations \eqref{eq:YuleWalker} (applied to the original AR($p$) process), we have
\begin{align}
    \Cn{p+1}\begin{bmatrix}\bbba\\1\end{bmatrix}=\begin{bmatrix}\bzero\\\sigma_W^2\end{bmatrix}
    ,
\end{align} 
or equivalently, 
\begin{align}
    \begin{bmatrix}\bq\\q\end{bmatrix} &=
    \Cn{p+1}^{-1}\begin{bmatrix}\bzero\\ 1\end{bmatrix}
    = \sigma_W^{-2}\begin{bmatrix}\bbba\\1\end{bmatrix},
\end{align}
so we identify:
\begin{align}
    \left[\Cn{p+1}^{-1}\right]_{p+1,p+1}&=q=\sigma_W^{-2}, 
 \\ \left[\Cn{p+1}^{-2}\tbc_{p+1}\right]_{p+1}&=-\left[\Cn{p+1}^{-1}\begin{bmatrix}\bba\ 0\end{bmatrix}\right]_{p+1}=-\bq^T\bba=-\sigma_W^{-2}(\bbba^T\bba),
 \\ \left[\Cn{p+1}^{-2}\right]_{p+1,p+1}&=\bq^T\bq+q^2=\sigma_W^{-4}(\bbba^T\bbba+1)=\sigma_W^{-4}(\ba^T\ba).
\end{align}
By substituting in \eqref{eq:Lpp1}, we get (using $c_0+\bba^T\bbc_p=\ba^T\bc_p$ and $1+\bba^T\bba=\ba^T\ba$)
\begin{align}
    \LMMSEtwoP{p+1}=\frac{\ba^T\bc_p-\sigma_W^{-2}\mu^2}
    {\ba^T\ba-2(\bbba^T\bba)\sigma_W^{-2}\mu+(\ba^T\ba)\sigma_W^{-4}\mu^2}.
\end{align}
By reparametrizing $\tilde{\mu}\triangleq\sigma_W^{-2}\cdot\mu$, we obtain
\begin{align}
    \LMMSEtwoP{p+1} &= \frac{\ba^T\bc_p-\sigma_W^2\cdot\tilde{\mu}^2}
    {\ba^T\ba-2(\bbba^T\bba)\tilde{\mu}+(\ba^T\ba)\tilde{\mu}^2}
 \\ &= \frac{\ba^T\bc_p(1-\tilde{\mu}^2)}{\ba^T\ba(\tilde{\mu}^2-2\alpha^{-1}\tilde{\mu}+\tilde{\mu}^2)}
\end{align}
(where we have used $\sigma_W^2=\ba^T\bc_p$ and where 
\begin{align}
    \alpha\triangleq\frac{\ba^T\ba}{\bbba^T\bba}=\frac{\sum_{\ell=0}^p a_\ell^2}{\sum_{\ell=1}^p a_\ell a_{p+1-\ell}}
\end{align}
as defined in the Theorem), to be maximized w.r.t.\ $\tilde{\mu}$. Straightforward calculations show that the maximizing solution is given by
\begin{equation}
    \tilde{\mu}=\alpha-\sign(\alpha)\sqrt{\alpha^2-1},
\end{equation}
which concludes the proof by retrieving $\mu=\tilde{\mu}\cdot\sigma_W^2.$


\section{Proof of \thmref{thm:mintin-C}}
\label{app:mintin-C}

    Using the notations defined in 
    \appref{app:mintin-AR}, 
    we partition $\Cn{p+2}$ and its inverse as follows, using 
    \lemref{lem:InverseMatrix:Schur}.
    \begin{align}
        \Cn{p+2}=\begin{bmatrix}\Cn{p+1} & \bbbc_{p+1}\\ \bbbc_{p+1}^T & c_0\end{bmatrix},
    \end{align}
    or equivalently, 
    \begin{align}
        \Cn{p+2}^{-1}=\begin{bmatrix}\bQ^{-1} & \bq \\ \bq^T & q\end{bmatrix},
    \end{align}
    where
    \begin{align}
        \bQ=\Cn{p+1}-\frac{1}{c_0}\bbbc_{p+1}\bbbc_{p+1}^T ,
    \end{align} 
    and 
    \begin{align} 
        q=\frac{1}{c_0^2}\left(c_0+\bbbc_{p+1}^T\bQ^{-1}\bbbc_{p+1}\right)
    \end{align}
    (the value of $\bq$ is irrelevant here). Therefore, $\trace{\Cn{p+2}^{-1}}=\trace{\bQ^{-1}}+q$. We use  \eqref{eq:matrix-inversion-lemma} 
    (which reduces to the Sherman--Morrison formula  \cite[Ch.~0.7]{HornJohnsonBook:MatrixAnalysis} in this case) to obtain $\bQ^{-1}$: 
    \begin{align}
        \bQ^{-1}=\Cn{p+1}^{-1}+\frac{(\Cn{p+1}^{-1}\bbbc_{p+1})(\Cn{p+1}^{-1}\bbbc_{p+1})^T}{c_0-\bbbc_{p+1}^T\Cn{p+1}^{-1}\bbbc_{p+1}}.
    \end{align}
    Note also that
    \begin{align}
        \bbbc_{p+1}^T\bQ^{-1}\bbbc_{p+1}=\bbbc_{p+1}^T\Cn{p+1}^{-1}\bbbc_{p+1}+\frac{(\bbbc_{p+1}^T\Cn{p+1}^{-1}\bbbc_{p+1})^2}{c_0-\bbbc_{p+1}^T\Cn{p+1}^{-1}\bbbc_{p+1}},
    \end{align}
    so that we can now express the full Tin as
    \begin{subequations}
    \noeqref{eq:trCpp2:1,eq:trCpp2:2,eq:trCpp2:3,eq:trCpp2:4} 
    \begin{align}
        &\trace{\Cn{p+2}^{-1}}=\trace{\Cn{p+1}^{-1}}+\frac{(\Cn{p+1}^{-1}\bbbc_{p+1})^T(\Cn{p+1}^{-1}\bbbc_{p+1})}{c_0-\bbbc_{p+1}^T\Cn{p+1}^{-1}\bbbc_{p+1}}
     \\ &\quad\qquad\qquad\qquad 
        + \frac{1}{c_0^2}\left(c_0+\bbbc_{p+1}^T\bQ^{-1}\bbbc_{p+1}\right)
     \label{eq:trCpp2:2}
     \\ &=\trace{\Cn{p+1}^{-1}}+\frac{\bbbc_{p+1}^T\Cn{p+1}^{-2}\bbbc_{p+1}}{c_0-\bbbc_{p+1}^T\Cn{p+1}^{-1}\bbbc_{p+1}}
     \\ &\quad + \frac{1}{c_0^2}\left(c_0+\bbbc_{p+1}^T\Cn{p+1}^{-1}\bbbc_{p+1}+\frac{(\bbbc_{p+1}^T\Cn{p+1}^{-1}\bbbc_{p+1})^2}{c_0-\bbbc_{p+1}^T\Cn{p+1}^{-1}\bbbc_{p+1}}\right)
     \label{eq:trCpp2:4}
     \\ &=\trace{\Cn{p+1}^{-1}}+\frac{1+\bbbc_{p+1}^T\Cn{p+1}^{-2}\bbbc_{p+1}}{c_0-\bbbc_{p+1}^T\Cn{p+1}^{-1}\bbbc_{p+1}}.
     \label{eq:trCpp2}
   \end{align}
   \end{subequations}
    Noting that, due to the bisymmetry\footnote{A square matrix $\bA$ is bisymmetric if it is symmetric about both of its main diagonals, satisfying $\bA=\bA^T=\bJ\bA\bJ$, where $\bJ$ denotes the unit anti-diagonal (reversal) matrix; the inverse of a bisymmetric matrix is bisymmetric, and the product of a bisymmetric matrix with itself is bisymmetric.} of $\Cn{p+1}$ and its inverse, we have $\bbbc_{p+1}^T\Cn{p+1}^{-i}\bbbc_{p+1}=\bbc_{p+1}^T\Cn{p+1}^{-i}\bbc_{p+1}$ ($i=1,2$), we identify the last term in \eqref{eq:trCpp2} (affected by $\bbc_{p+1}$) as the reciprocal of $\LMMSEtwoP{p+1}$ \eqref{eq:Lpp1a}. Therefore, the vector $\bbc_{p+1}$ minimizing $\trace{\Cn{p+2}^{-1}}$ (subject to complying with the given covariance values) is the same $\bbc_{p+1}$ that maximizes $\LMMSEtwoP{p+1}$ under the same constraint, and therefore the optimal solution for the only free variable $c_{p+1}$ in that vector is the same as in \thmref{thm:mintin-AR}.



\end{document}